\documentclass[acmsmall]{acmart}
\usepackage{xspace}
\usepackage{enumerate}
\usepackage[shortlabels]{enumitem}
\usepackage{multirow}
\usepackage{changepage}
\usepackage{subcaption}

\AtBeginDocument{%
  \providecommand\BibTeX{{%
    \normalfont B\kern-0.5em{\scshape i\kern-0.25em b}\kern-0.8em\TeX}}}

\setcopyright{acmlicensed}
\acmJournal{PACMHCI}
\acmYear{2022} \acmVolume{6} \acmNumber{CSCW2} \acmArticle{545} \acmMonth{11} \acmPrice{15.00}\acmDOI{10.1145/3555603}

\received{January 2022}
\received[revised]{April 2022}
\received[accepted]{August 2022}

\newcommand{\cut}[1]{}

\newcommand{\rredit}[1]{{\textcolor{purple}{#1}}}
\newcommand{\rrcomment}[1]{{\textcolor{navy}{#1}}}
\newcommand{\rrstat}[1]{{\textcolor{burgundy}{#1}}}
\newcommand{\rrdelete}[1]{{\textcolor{gray}{\sout{#1}}}}

\newcommand{\rrdeletesec}[1]{{\color{gray}{\section*{\sout{#1}}}}}
\newcommand{\rrdeletesubsec}[1]{{\color{gray}{\subsection*{\sout{#1}}}}}
\newif\ifmarkup
\markupfalse
\ifmarkup
\else
\renewcommand{\rredit}[1]{#1}
\renewcommand{\rrcomment}[1]{}
\renewcommand{\rrstat}[1]{}
\renewcommand{\rrdelete}[1]{}

\renewcommand{\rrdeletesec}[1]{}
\renewcommand{\rrdeletesubsec}[1]{}
\fi

\newcommand{\mredit}[1]{{\textcolor{purple}{#1}}}
\newif\ifmrmarkup
\mrmarkupfalse
\ifmrmarkup
\else
\renewcommand{\mredit}[1]{#1}
\fi

\begin{document}

\title{Thread With Caution: Proactively Helping Users Assess and Deescalate Tension in Their Online Discussions}
\renewcommand{\shorttitle}{Thread With Caution}

\author{Jonathan P. Chang}
\email{jpc362@cornell.edu}
\affiliation{%
  \institution{Cornell University}
  \city{Ithaca}
  \state{NY}
  \country{USA}
  \postcode{14850}
}
\author{Charlotte Schluger}
\email{jes543@cornell.edu}
\affiliation{%
  \institution{Cornell University}
  \city{Ithaca}
  \state{NY}
  \country{USA}
  \postcode{14850}
}
\author{Cristian Danescu-Niculescu-Mizil}
\email{cristian@cs.cornell.edu}
\affiliation{%
  \institution{Cornell University}
  \city{Ithaca}
  \state{NY}
  \country{USA}
  \postcode{14850}
}

\renewcommand{\shortauthors}{Jonathan P. Chang, Charlotte Schluger, \& Cristian Danescu-Niculescu-Mizil}

\begin{abstract}
Incivility remains a major challenge for online discussion platforms, to such an extent that even conversations between well-intentioned users can often derail into 
uncivil behavior.   
Traditionally, platforms have relied on moderators to---with or without algorithmic assistance---take corrective actions such as removing comments or banning users.
In this work we propose 
a complementary
paradigm that directly empowers users by proactively enhancing their awareness about  
existing tension in the conversation they are engaging in
 and actively guides them as they are drafting their replies to avoid further escalation.

\rredit{As a proof of concept for} this paradigm, we design an algorithmic tool 
that provides such proactive information directly to users,
and conduct a user study in a popular discussion platform.  
Through a mixed methods approach combining surveys with a \rct,%
\rredit{
we uncover qualitative and quantitative insights regarding how the participants utilize and react to this information. 
Most participants report finding this proactive paradigm valuable, noting that it helps them to identify tension that they may have otherwise missed and prompts them to further reflect on
 their own replies
and to revise them. 
These effects are corroborated by a comparison of how the participants draft their reply when our tool warns them that their conversation is at risk of derailing into uncivil behavior versus in a control condition where the tool is disabled.
These preliminary findings highlight the potential of this user-centered paradigm and point to concrete directions for 
future implementations.
}

\end{abstract}



\newcommand{\xhdr}[1]{{\noindent\bfseries #1.}}
\newcommand{\convowizard}{ConvoWizard\xspace}
\newcommand{\ourparadigm}{advance warning\xspace}
\newcommand{\contextfeed}{Context Summary\xspace}
\newcommand{\replyfeed}{Reply Summary\xspace}
\newcommand{\treatment}{Treatment\xspace}
\newcommand{\control}{Control\xspace}
\newcommand{\atrisk}{at-risk\xspace}
\newcommand{\Atrisk}{At-risk\xspace}
\newcommand{\notatrisk}{not-at-risk\xspace}
\newcommand{\Notatrisk}{Not-at-risk\xspace}
\newcommand{\riskawareness}{risk awareness\xspace}
\newcommand{\cmv}{ChangeMyView\xspace}
\newcommand{\rct}{randomized controlled experiment\xspace}
\newcommand{\participant}[1]{\textbf{P{#1}}\xspace}

\maketitle

\section{Introduction}
\label{sec:intro}

Incivility remains an important issue in online discussion platforms \cite{chandrasekharan_you_2017}, hindering the exchange of ideas \cite{arazy_stay_2013} and taking a significant emotional toll on the participants \cite{ashktorab_designing_2016,jhaver_view_2018}.
Traditionally, platforms attempt to address this problem through reactive moderation, in which volunteers from within the community \cite{seering_reconsidering_2020} or professionals employed by the platform operator \cite{gillespie_custodians_2018} aim to identify and remove ``bad actors'' and ``objectionable content''.
Substantial efforts are focusing on scaling up this paradigm through automation or algorithmic assistance, an enterprise which has proven to be both technically and ethically challenging \cite{grimmelmann_virtues_2015,gillespie_content_2020,katzenbach_ai_2021,gorwa_algorithmic_2020}.

This common paradigm, however, does not account for the fact that 
uncivil behavior in discussion platforms is not solely the product of ``bad actors''\rredit{---who are generally a minority within their communities \cite{kumar_community_2018}---}but can instead often emerge from 
\rredit{ordinary} users when they find themselves in particularly heated or tense situations \cite{cheng_anyone_2017}.
In fact, 
in many settings%
\rredit{
the vast majority of individuals on a platform are \emph{well-intentioned}, in the sense that their purpose for being on the platform is simply to consume interesting content and engage in good faith with other community members \cite{srinivasan_content_2019,gilbert_i_2020,weld_what_2022}.
}
Starting from this viewpoint, in this work we propose a complementary paradigm that directly empowers such well-intentioned users to proactively avoid escalating tense situations.
It does so by informing them when the conversation they are engaging in is at risk of derailing into incivility, i.e., enhancing their \emph{risk awareness}.

To test whether this 
proactive 
\riskawareness paradigm is feasible in a real world setting---and thus avoid artifacts of laboratory or crowdsourced studies \cite{taylor_accountability_2019,reinecke_labinthewild_2015,goodman_data_2013}---we conduct a user study in a popular online discussion platform, \cmv. 
Executing this real-world user study requires us to engage with several core challenges: not only technical, but also practical and ethical.

From a technical perspective, we need to build a system that can both automatically assess the risk of derailment for ongoing discussions in real time and inform the participants about this risk and about the potential impact of their responses \emph{as they are drafting them}.
To this end, we develop a prototype tool, which we call \convowizard, consisting of a backend algorithmic scoring system powered by a recent natural language processing methodology---conversational forecasting~\cite{zhang_conversations_2018,liu_forecasting_2018,chang_trouble_2019}---and a frontend browser plugin that serves the resulting information to users directly on the \cmv webpage (see the Video Figure).\footnote{A link to the Video Figure can be found at \url{https://www.cs.cornell.edu/~cristian/Thread_With_Caution.html}.}

From an ethical and practical perspective, turning regular platform users into volunteers for a scientific study 
requires
a design that puts their needs and well-being at the fore.  
Thus, in designing and conducting this user study, we adopted a community collaboration model which took direct input from \cmv community leaders.
Additionally, we used a two-phase study design, starting with a larger phase in which we sought feedback from the participants after using the fully functional tool for one month, and continuing with a second phase in which we implemented a within-participant \rct lasting two months.

The results of the user study suggest that the risk awareness paradigm has the potential to improve online discourse and motivate further research 
in this direction.  
In exit surveys, the majority of participants report%
\rredit{
that they found \convowizard helpful for identifying tense situations, 
with the tool
both
\emph{supplementing}
their
intuitions---catching types of tension that 
they
may not have known to look for---and 
\emph{activating} their existing intuitions---reminding them to be on the lookout for tension in situations where they may not have been paying attention. 
Most participants 
also
 report that this additional awareness of risk helped them avoid fights and kept them from posting comments they would have regretted later.
}

\rredit{
Combining 
 feedback from participants with quantitative analysis of the 
  data from the \rct
 offers a glimpse into concrete steps participants took in response to this increased awareness. 
First, participants report that seeing a warning from \convowizard led them to 
reflect more
 on the tension in the conversation and how their reply might affect it. 
This effect is echoed in the randomized controlled experiment results: when users are warned that a conversation they are participating in is at risk of future incivility, they spend 9\% more time on average drafting their comment compared to the control condition where they are not warned
of this risk.
Beyond reflecting on tension, participants further report that they go on to revise their draft reply using \convowizard as a guide
to reduce the risk of derailment.
This effect is again echoed in the experimental results: when users are warned about 
an
existing risk they edit their reply in a way that tends to gradually decrease this risk, whereas in the control condition where they are not warned, they tend to escalate the risk. 
While the observed effects are 
small and 
limited by the scale of our study, they nonetheless combine with our qualitative observations to offer a promising initial indicator that directly empowering well-intentioned users with additional awareness about risk of derailment is a feasible complement to existing moderation practices, with potential to improve online discourse. 
This establishes a groundwork for future studies by highlighting concrete directions 
for future implementations of this paradigm.
}

In summary, in this paper we:

\begin{itemize}
    \item propose a new paradigm that empowers well-intentioned users to assess and address the risk of incivility in the conversations they participate in;
    \item develop a fully functional tool that implements this paradigm in a popular discussion community; and
    \item design and conduct a user study, in collaboration with the moderators of this community, to evaluate the feasibility and potential of this new paradigm.
\end{itemize}

\section{Background and Related Work}
\label{sec:background}
Every online discussion platform must eventually reckon with the problem of incivility and other undesirable behaviors, leading to \citeauthor{gillespie_custodians_2018}'s assertion that ``all platforms moderate'' \cite{gillespie_custodians_2018}.
Yet the \emph{specific} practices for tackling this problem vary greatly across different platforms, and in turn a rich body of prior literature on moderation has sought to categorize and compare these different strategies.
To better contextualize our work within the landscape of moderation, we first synthesize prior work's categorization of moderation practices, identifying two key axes of design: \emph{who} does the work of moderation and \emph{when} this work happens.
Then, we explain where our proposed paradigm falls within this typology.

\subsection{Actors Involved in Moderation: Who Does the Work?}
\label{sec:actors}
Today, many online platforms take a \emph{platform-driven} approach to moderation, where platform operators directly employ or contract workers to review potentially objectionable content and remove it if needed \cite{gillespie_custodians_2018}.
This is arguably the model of moderation that the lay audience is most familiar with, as it has been adopted by the most prominent platforms such as Facebook and Twitter, and has been a driving force behind high-profile moderation cases such as Reddit's 2015 mass ban of hate communities \cite{chandrasekharan_you_2017}.
However, this strategy also suffers from key weaknesses.
Chief among these is the problem of \emph{scale}: the large amount of content being generated on major online platforms makes it infeasible for moderators to handle all content needing review in a timely manner \cite{gillespie_content_2020}, and results in a high workload and stress for the moderators \cite{roberts_behind_2014}.

Though the platform-driven approach may be dominant in today's Web, its ascendancy was by no means a foregone conclusion: early online communities, with their decentralized ethos, tended to instead prefer a bottom-up, \emph{community-driven} model \cite{dibbell_rape_2005,lampe_slashdot_2004}.
As of late, community-driven moderation has seen a renewed surge in interest in light of the shortcomings of platform-driven moderation \cite{brewer_inclusion_2020,seering_reconsidering_2020}, and it remains the method of choice in smaller, interest-specific communities---for example, Twitch livestream communities \cite{lo_when_2018,cai_what_2019} and the topical groups on Reddit known as ``subreddits'' \cite{dosono_moderation_2019,chandrasekharan_internets_2018,gilbert_i_2020}.
Community-driven moderation practices can be further subdivided as roughly falling into two categories: volunteer moderators and end-user tools.

One common approach to community-driven moderation mimics the centralized model of platform-driven moderation, granting the authority to review and remove content to a core group of \emph{volunteer moderators}, who are not platform employees but rather regular community members who have stepped up to the task \cite{dosono_moderation_2019,geiger_work_2010,lo_when_2018,seering_reconsidering_2020,wohn_volunteer_2019}.
While volunteer moderators are conceptually similar to platform-employed moderators in terms of their administrative powers and workflow, their status as actual members of the communities they moderate can be a unique advantage: they may receive a higher level of trust and connection from the community, unlike platform-employed moderators who are seen as outsiders \cite{seering_reconsidering_2020}, and their inside knowledge of community norms and dynamics can help them negotiate harder, more nuanced disputes~\cite{turnbull_thats_2018,chandrasekharan_internets_2018}.
On the other hand, like their platform-employed counterparts, volunteer moderators face the problem of scale, and the resulting problems of overwork and stress are exacerbated by the fact that volunteer moderators are doing this work in their free time, not as their full-time job \cite{dosono_moderation_2019,wohn_volunteer_2019}.

As such, online communities have sought strategies to mitigate the problem of uncivil behavior outside the framework of centralized moderation, thereby decreasing the burden on moderators.
This has led to a second family of community-driven moderation strategies: \emph{end-user tools}.
In contrast to the previously described centralized strategies, end-user tools are accessible to \emph{all} community members, distributing the work of content moderation across the entire community~\cite{kiesler_regulating_2012,seering_reconsidering_2020}.
While the risk of misuse necessarily implies that end-user tools cannot be as authoritative as moderators' tools (e.g., end users should not have the ability to remove someone else's content), platforms have managed to innovate various softer approaches that have met with some success.
A particularly common end-user tool is the ability to \emph{vote} on whether a piece of content constitutes a valuable contribution to the community; content that receives too many negative votes can then be automatically de-prioritized or hidden \cite{lampe_slashdot_2004,chandrasekharan_internets_2018,mamykina_design_2011}.
An even softer end-user tool is the personalized \emph{blocklist} \cite{geiger_bot-based_2016,jhaver_online_2018}, which shifts the goal from removing objectionable content from the platform to simply removing it from an individual user's feed.

\subsection{When Does Moderation Work Happen?}
While the moderation strategies we have described so far are quite varied, they all have one thing in common: they are designed to deal with uncivil content that has \emph{already} been created.
While dealing with uncivil content after-the-fact is better than doing nothing, some have argued that a more effective way to protect online communities from harm is to reduce the amount of uncivil content that gets created in the first place \cite{kiesler_regulating_2012,grimmelmann_virtues_2015}.
Strategies designed to achieve this have been referred to as \emph{proactive} moderation \cite{lo_when_2018,seering_reconsidering_2020,seering_shaping_2017,cai_moderation_2021}, as a contrast to the previously described after-the-fact strategies that are referred to as \emph{reactive} moderation.

In current practice, proactive moderation is largely the realm of volunteer moderators, whose combination of authority and connection to the community put them in a unique position to engage in social strategies to guide user behavior towards healthier interactions \cite{seering_metaphors_2020,seering_moderator_2019}.
Such strategies include educating users about the community's rules \cite{seering_moderator_2019,cai_what_2019}, publicly modeling good behavior \cite{jagannath_we_2020,seering_shaping_2017}, or mediating disputes before they can get out of hand \cite{billings_understanding_2010}.
In interviews, volunteer moderators have indicated that they see this as just another part of their job, describing the work using metaphors like ``teacher'' and ``facilitator'' \cite{seering_metaphors_2020}.

\xhdr{Where our paradigm fits: User-facing proactive interventions} 
The proactive strategies we have discussed thus far involve moderator actions.
But as we have previously seen, when it comes to \emph{reactive} strategies, there is room for end users to play a role alongside moderators in the broader landscape of moderation.
Does the same hold true for proactive strategies?

This question has driven a recently emerging line of research that looks at how platform design could directly steer end users towards more prosocial behaviors, in the form of user-facing \emph{interventions}.
Various intervention strategies have been attempted: some work explicitly encourages users to take particular actions, such as reflecting more deeply on comments they have read \cite{kriplean_is_2012,kriplean_supporting_2012}; others operate on a more subconscious level, by asking users to complete tasks prior to commenting that might prime them to be more prosocial \cite{seering_designing_2019,taylor_accountability_2019}; and still others simply aim to provide users with additional information, such as the fact that the other user they are engaging with is new to the community \cite{halfaker_nice:_2011}, in hopes that this additional information might influence their subsequent behavior.

While experiments with these interventions have shown success, the authors of one such system, \citeauthor{taylor_accountability_2019} \cite{taylor_accountability_2019}, caution that their implementation (and others like it) suffer from a key vulnerability that might limit their effectiveness in the real world: 
they are \emph{static}, in the sense that they are globally applied across all of a user's interactions.
The problem is that not every interaction requires an intervention; in most interactions people are already behaving civilly.
Though at first this seems at most a minor annoyance, \citeauthor{taylor_accountability_2019} reason that because that peoples' capacity for empathy is finite, static interventions might only work in a limited lab setting---if users were seeing the intervention all the time in their everyday social media usage, they might get overwhelmed and just tune it out.
A similar ``attrition'' effect, where static interventions lose effectiveness over time when deployed at scale, has been observed in work on interventions in other fields \cite{krebs_meta-analysis_2010,kovacs_rotating_2018,collins_social_2014}.
Thus, \citeauthor{taylor_accountability_2019} argue, making proactive interventions effective at scale requires a \emph{dynamic} approach of ``targeting design interventions just in time for the individuals who need them.''

Our present work adds to the ongoing research on proactive intervention design by exploring \citeauthor{taylor_accountability_2019}'s proposal of targeted, just-in-time interventions.
We are inspired by a recent algorithmic development, the emergence of \emph{conversational forecasting} algorithms~(Section~\ref{sec:craft}) that could provide the technical backbone for \citeauthor{taylor_accountability_2019}'s proposal, automatically identifying interactions in need of intervention by detecting \emph{rising tension} that could lead to incivility in the future.
We use this technology to build our own implementation of a proactive intervention system (Section~\ref{sec:prototype_tool}), and heeding \citeauthor{taylor_accountability_2019}'s caution 
about the limitations of lab studies, 
we describe and execute a plan for evaluating the intervention in a real-world setting (Section \ref{sec:study_design}).

\section{Methods}
\label{sec:method}

To evaluate the feasibility of our proposed \riskawareness paradigm
 we develop a prototype tool that implements it, \convowizard (Section \ref{sec:prototype_tool}), and gather both qualitative feedback and quantitative usage data through an
IRB-approved
user study (Section \ref{sec:study_design}).
Heeding \citeauthor{taylor_accountability_2019}'s warnings of the potential inadequacies of 
laboratory studies
in understanding the impact of prosocial interventions,
we design our user study 
to capture how regular users might be affected by the intervention in their everyday online interactions.
Prior work throughout the HCI and CSCW space has argued that achieving this goal requires going outside the laboratory and testing the intervention in a real world setting, or ``in the wild'' \cite{reinecke_labinthewild_2015,consolvo_activity_2008,brown_into_2011,mottelson_virtual_2017}.
Following this line of work, we set up our user study to involve real users in an actual social media community, namely the Reddit debate forum \cmv.
However, the real-world setting also introduces a host of technical, practical, and ethical challenges, which end up shaping the design of our study:

\noindent\textbf{Technical challenge}: How can we 
provide users with real-time information about the risk of real online conversations?
In a laboratory setting, the researchers would have full control over both the conversations that get shown (which would enable them to pre-annotate the risk of each conversation) and the UI of the simulated platform (which would enable them to easily add the risk information as an additional UI element).
By contrast, real online conversations take place on established platforms that we lack control over.
In Section \ref{sec:prototype_tool}, we explain the technical approach we take to tackling this problem, developing a browser extension that uses established techniques to read the content of conversations taking place on Reddit, algorithmically score the risk level of that content in real time, and extend the Reddit UI with additional elements that can be used to display interventions based on the score.

\noindent\textbf{Practical challenge}: How can we convince everyday users of online platforms to use our tool as part of their regular activity?
In particular, since we implement our interventions via a browser extension, participants need to be willing to not only install the software but also keep it enabled for the full duration of the study.
Therefore, the tool needs to provide real value to the user in addition to supporting the research.  
In section \ref{sec:experimental_conditions}, we explain how we set up the experimental conditions in order to combine these goals.

\noindent\textbf{Ethical challenge}: Algorithmic systems can produce flawed or biased judgments \cite{davidson_automated_2017,duarte_mixed_2018}, and harm can occur if such flawed judgments are used as the basis of real-world actions.
In the specific context of our study, this could take the form of our tool providing wrong estimates of risk to users, which might cause them to make bad decisions.
Because our study is taking place in real online discussions, the potential harm is not just limited to the 
study participants
themselves, but to other 
users in the discussion, and perhaps even the broader community.
This 
danger
carries a clear ethical implication: because the community shoulders the 
potential harms
arising from flaws or misuse of our technology, the community should be consulted and involved in the running of the study.
This conclusion leads us to develop our study as a \emph{community collaboration}, done as a joint endeavor with the moderators of \cmv.
In Section \ref{sec:community}, we explain this approach in more detail.

\subsection{Technical Design: The \convowizard Tool}
\label{sec:prototype_tool}
To address the technical challenge of presenting users with 
advance
notification of how their comments may affect a conversation, we build \convowizard: a prototype tool that is  designed to assess the risk of conversations in real time and deliver this information to the user. 
\convowizard is comprised of two parts: (1) a browser extension we distributed to participants in the study which extracts data about the conversations they engage with on \cmv, collects data about their in-progress drafts, and displays UI interventions; and (2) a backend server which runs a machine learning model in real time to predict the trajectory of ongoing conversations, relays this information to the browser extension, and logs data for subsequent analysis.
\subsubsection{Underlying machine learning model: The conversational forecasting algorithm}
\label{sec:craft}

To automatically estimate the risk of future incivility in conversations, \convowizard leverages a recent Natural Language Processing paradigm, \textit{conversational forecasting}, which trains models to predict future conversational outcomes based on the current state of the conversation \cite{zhang_conversations_2018,liu_forecasting_2018,chang_trouble_2019}.
Specifically, \convowizard uses CRAFT, a conversational forecasting model that was trained to forecast the future occurrence of uncivil behavior in a conversation \cite{chang_trouble_2019}.
An appealing aspect of CRAFT for our study is that it has previously been trained and evaluated on \cmv, using ``natural'' labels that came from \cmv moderator actions, specifically removals of comments that violated \cmv's rules against uncivil behavior.\footnote{We use the publicly available \cmv CRAFT model from the ConvoKit package (\url{https://convokit.cornell.edu}).}
Formally, given a conversation $C = \lbrace c_1, c_2, \dots, c_n \rbrace$ represented as a series of comments in reply-to order, CRAFT predicts the likelihood 
that
the next comment $c_{n+1}$ 
will contain uncivil behavior. In other words, $\text{CRAFT}(C) = p(\text{isUncivil}(c_{n+1}))$. In this way, given the current state of a conversation, CRAFT can predict the 
risk
 that the next comment that gets posted will exhibit uncivil behavior. Moreover, CRAFT is an online model: when a new comment $c_{n+1}$ is eventually added to the discussion, CRAFT can compute an updated prediction for the chance of future antisocial behavior by including the new comment for consideration: $p(\text{isUncivil}(c_{n+2})) = \text{CRAFT}(\lbrace c_1, c_2, \dots, c_n, c_{n+1} \rbrace)$.

\begin{figure}
    \begin{subfigure}[b]{0.75\textwidth}
        \includegraphics[width=\textwidth]{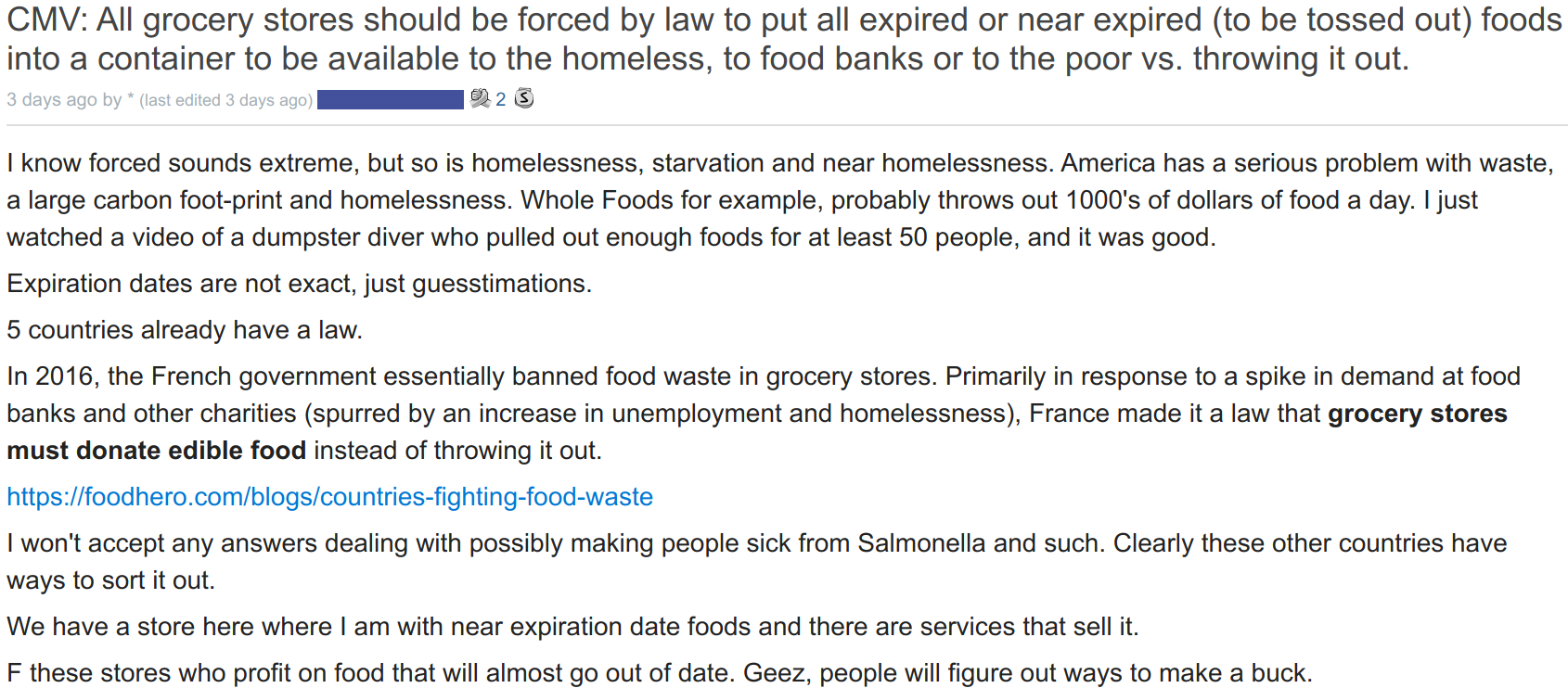}
        \caption{}
        \label{fig:context_post}
    \end{subfigure}
    \begin{subfigure}[b]{0.475\textwidth}
        \includegraphics[width=\textwidth]{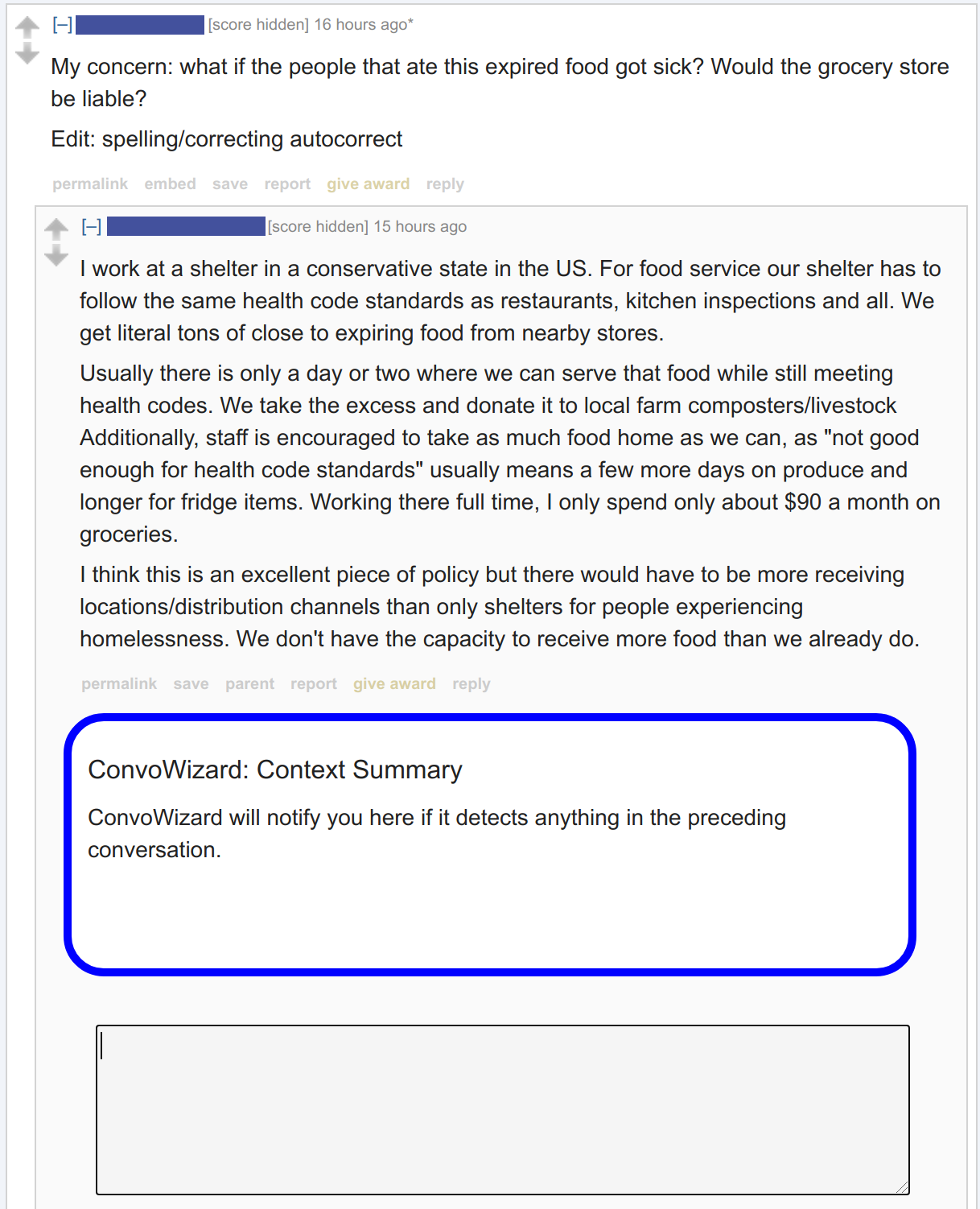}
        \caption{}
        \label{fig:context_clean}
    \end{subfigure}
    \begin{subfigure}[b]{0.475\textwidth}
        \includegraphics[width=\textwidth]{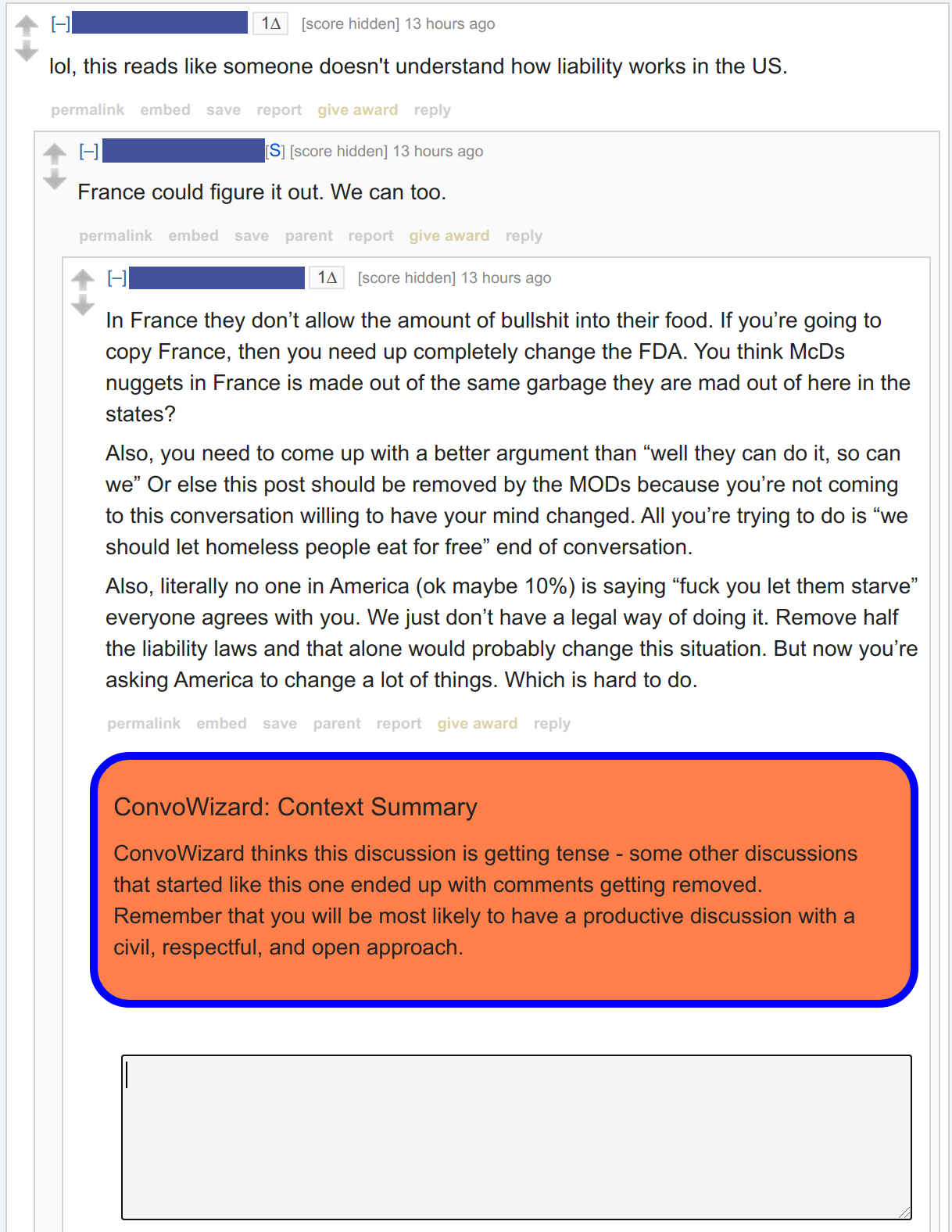}
        \caption{}
        \label{fig:context_tense}
    \end{subfigure}
    \caption{The \contextfeed feature of \convowizard provides information about whether the conversation 
    the user is joining is at risk of turning uncivil in the future.
    (b) When no risk is detected, the \contextfeed displays a neutral message on a blank background. (c) When risk is detected, the \contextfeed displays a warning message displayed on a red background, with deeper shades of red indicating higher risk. Note that both examples come from the same discussion thread; for reference, the post that started the thread is shown in (a). 
    }
    \label{fig:context_summary}
\end{figure}

\begin{figure}
    \begin{subfigure}[b]{0.475\textwidth}
        \includegraphics[width=\textwidth]{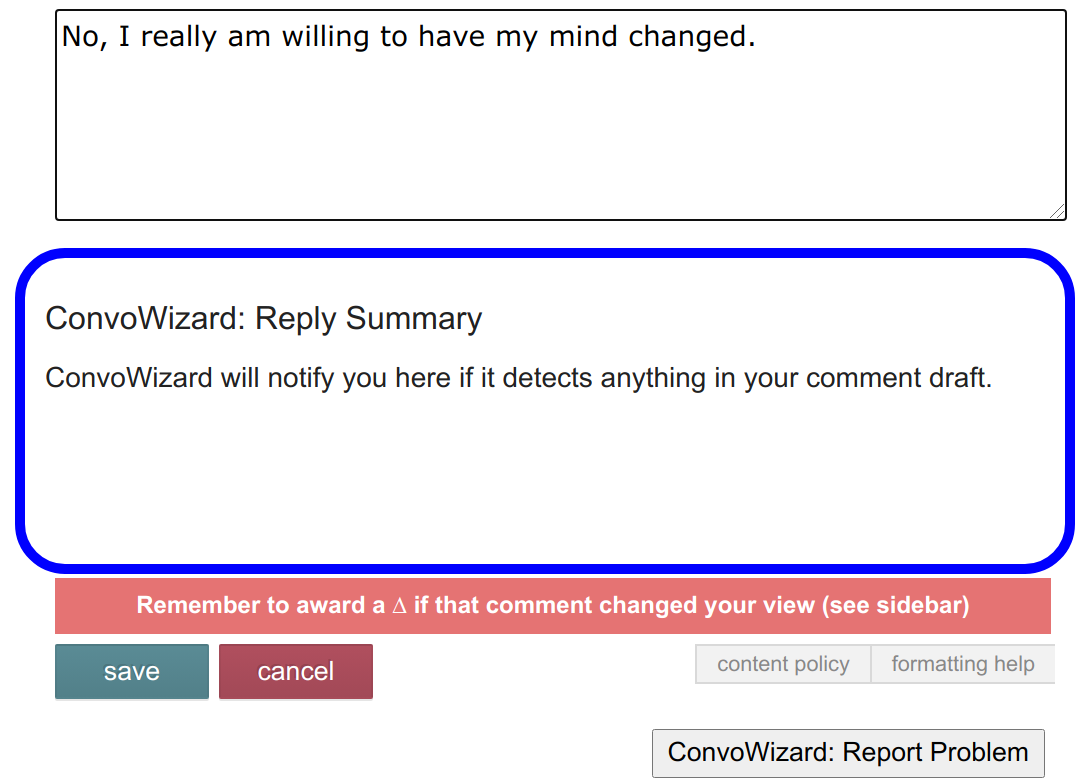}
        \caption{}
        \label{fig:reply_neutral}
    \end{subfigure}
    \begin{subfigure}[b]{0.475\textwidth}
        \includegraphics[width=\textwidth]{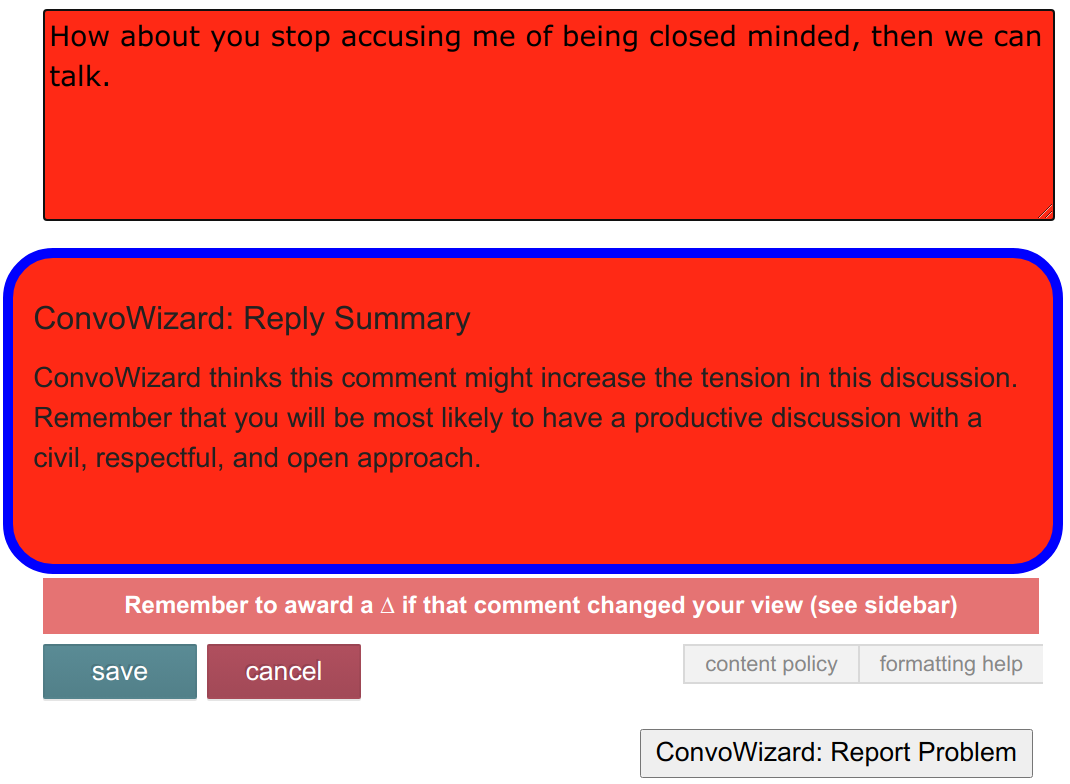}
        \caption{}
        \label{fig:reply_tense}
    \end{subfigure}
    \begin{subfigure}[b]{0.475\textwidth}
        \includegraphics[width=\textwidth]{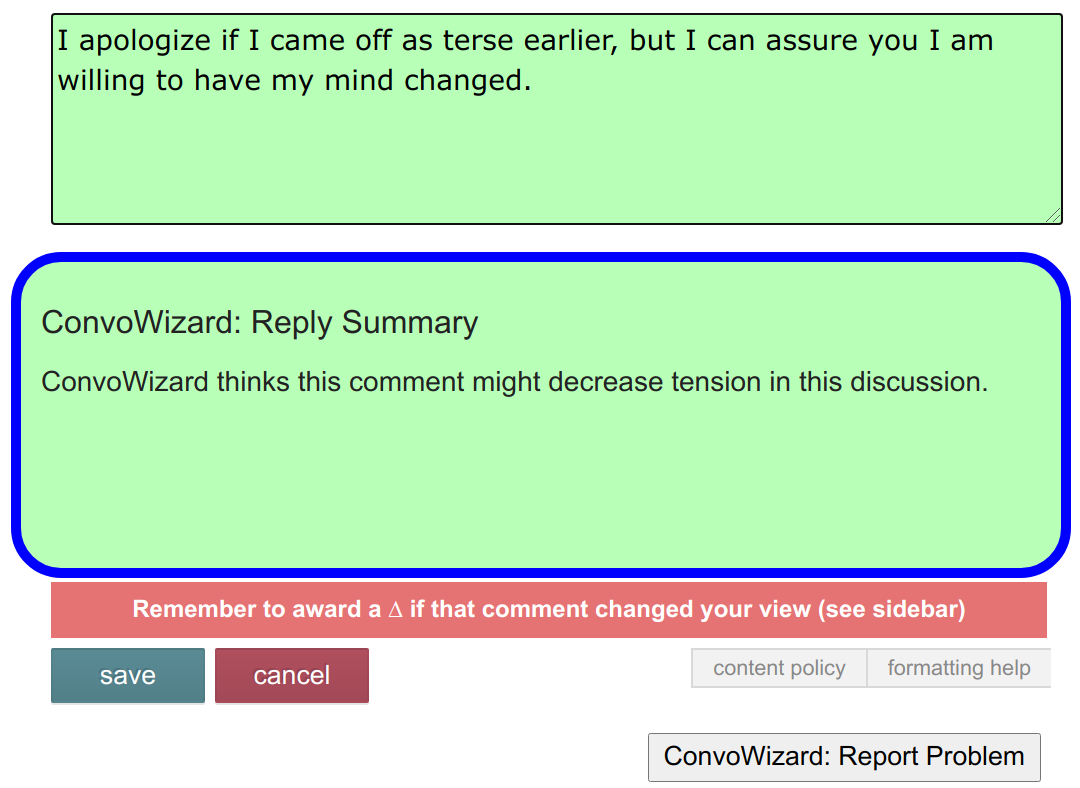}
        \caption{}
        \label{fig:reply_better}
    \end{subfigure}
    \caption{The \replyfeed provides information about what impact the user's in-progress draft reply might have on the risk of incivility. (a) If the risk score with the draft reply is the same as the risk score without the draft reply (within a margin of error), the \replyfeed displays a neutral message. (b) If the risk score increases, the \replyfeed displays a warning message with a red background, with deeper shades of red indicating higher resulting risk. (c) If the risk score decreases, the \replyfeed displays a message about decreased tension with a green background, with deeper shades of green indicating larger magnitudes of score decrease. (Note that all three examples shown are replies to the tense context from Figure \ref{fig:context_tense}; the preceding context is excluded for readability.)}
    \label{fig:reply_summary}
\end{figure}

\subsubsection{Frontend: the user facing extension}
\label{sec:frontend}

\convowizard's user-facing frontend is implemented as a Google Chrome extension which operates by reading and manipulating Reddit's browser-side HTML DOM,\footnote{Document Object Model, the browser's internal JavaScript-compatible representation of the web page.}
and therefore does \emph{not} require any access to the user's Reddit account.
The \convowizard extension activates whenever a user hits the "reply" button in the Reddit UI, indicating they are considering joining the discussion. 
As the user drafts their reply, \convowizard provides feedback directly inside the Reddit UI via DOM manipulation.
It specifically provides two types of feedback, referred to as the \contextfeed and the \replyfeed, which are each displayed in separate UI elements (demonstrated in the Video Figure).

The \textbf{\contextfeed} gives an estimate of how likely the conversation was to turn uncivil \emph{prior} to the user joining in.
To produce this estimate, the extension extracts the text of all preexisting comments $\lbrace c_1,\dots,c_n \rbrace$ in the conversation history from the DOM.
Then, it sends this information to the backend server (Section \ref{sec:backend}) which returns a CRAFT score $S_{context} = \text{CRAFT}(\lbrace c_1,\dots,c_n \rbrace)$; henceforth we refer to these scores as \emph{risk scores} to emphasize that in the context of \convowizard, CRAFT is being used as an estimate of the risk of future incivility. 
If $S_{context} > 0.55$,\footnote{The 0.55 threshold is what was recommended in the original CRAFT paper.} the \contextfeed displays a warning to the user that the conversation they are about to participate in is tense and might become uncivil in the future.
It also visually indicates this risk by changing its background color to a shade of red (scaling by risk score, such that higher scores produce redder colors).
This functionality is visualized in Figure \ref{fig:context_summary} and in the Video Figure.

Then, as the user drafts their reply, the \textbf{\replyfeed} provides real-time estimates of how the in-progress reply, if posted as-is, might impact the risk of 
the conversation turning uncivil in the future.
Every five seconds, the extension sends the current text of the in-progress reply, which we call $r(t)$ (where $t$ represents the current timestamp), to the \convowizard backend, which returns a risk score that was computed with this text included: $S_{r(t)} = \text{CRAFT}(\lbrace c_1,\dots,c_n,r(t) \rbrace)$.
The \replyfeed then determines what feedback to give by comparing $S_{context}$ and $S_{r(t)}$.\footnote{All comparisons apply a small noise threshold to prevent basing feedback on spurious variance in scores.}
If $S_{r(t)} > S_{context}$, the \replyfeed displays a warning that the in-progress reply might increase the tension in the conversation, and visually indicates this with a red background whose shade scales with $S_{r(t)}$.
On the other hand, if $S_{r(t)} < S_{context}$ and there was preexisting tension in the conversation (i.e., $S_{context} > 0.55$), the \replyfeed displays a message that the in-progress reply might decrease the tension, and visually indicates this with a green background whose shade scales with $S_{context} - S_{r(t)}$.\footnote{The ``decreasing tension'' intervention is only implemented for conversations with tension in the context because initial testers reported that it was confusing to hear about ``decreasing tension'' when there was no tension to begin with.}
This functionality is visualized in Figure \ref{fig:reply_summary} and in the Video Figure. 

\subsubsection{Backend server}
\label{sec:backend}
\convowizard also consists of a backend server component which is responsible for both running CRAFT to produce risk scores requested by the frontend, and logging the request data to produce a record of how users interacted with \convowizard.
Every time the backend receives a request from the frontend, it first runs CRAFT on the attached data to produce a risk score to return to the frontend, then it logs the request and response to a database.
Each logged request/response object includes the Reddit ID of the comment being replied to, the timestamp $t$ of the request, the generated risk score, and (for \replyfeed requests) the in-progress reply text $r(t)$.
Additionally, all requests that happened under the same reply action (i.e., the initial \contextfeed request that was sent when the user hit the ``reply'' button and all subsequent \replyfeed requests
until the reply is submitted or cancelled) are grouped together in the database as a single \emph{interaction}.
Knowing that a series of requests came from a single interaction allows us to subsequently analyze how users modified their drafts over time, as we will discuss in Section \ref{sec:findings}.

\subsubsection{Ethical considerations for technical design}
\label{sec:design_ethics}
As previously mentioned, the real-world setting of our study raises important ethical challenges.
While we primarily respond to these challenges through 
the
 design of the study, as we will discuss in Section \ref{sec:community}, there are also ethical implications for the design of the \convowizard tool itself.

First, there is the problem of \emph{misuse}:
\rredit{our \riskawareness paradigm is designed for well-intentioned users, who do not deliberately desire conflict} and are thus more likely to respond appropriately to warnings of potential future incivility.
By contrast, bad-faith trolls could respond to such warnings quite differently, for example purposely trying to write a comment that triggers a warning.
Thus, it is important to restrict access to \convowizard so that bad-faith trolls cannot easily get ahold of it.
To this end, \convowizard is programmed to be inoperable until it is ``activated'' using unique credentials that we assign to each participant in our study.
To prevent bad-faith trolls from circumventing this restriction by simply signing up for the study, we check the posting history of each user who signs up for our study, and prevent them from joining if they do not have an established history of participation on \cmv (as this might indicate that the account is a purpose-made ``sockpuppet'' \cite{kumar_army_2017} or an outsider seeking to ``brigade'' the subreddit \cite{georgakopoulou_making_2020}).

There is also the problem of \emph{errors} in \convowizard's algorithm-driven estimates of risk.
Algorithms that operate on human language, and especially on 
subjective
 aspects like civility, are far from perfect---they can fail to pick up on nuances of human behavior \cite{gillespie_content_2020,duarte_mixed_2018} and encode biases present in their training data \cite{davidson_automated_2017}.
But in the public consciousness, the capabilities and objectivity of algorithms are often overestimated \cite{bory_deep_2019,katzenbach_ai_2021}.
To address this we have crafted the messaging in and around the \convowizard tool to counteract 
such possible overestimation by users.
Throughout the instructions all study participants must read to set up \convowizard, we repeatedly remind them that \convowizard is an early prototype and may therefore make mistakes, and we encourage them to report any mistakes they notice.
Furthermore, the warning messages displayed in the \convowizard browser extension were specifically crafted to come across as \emph{informational} rather than \emph{prescriptive}---we avoided any wording that might imply the tool is advising users that they should (or should not) post their draft, as well as any language that might be associated with assigning blame.

The final wording, which frames \convowizard findings as simply the existence or nonexistence of ``tension'' in the conversation (see Figures \ref{fig:context_summary} and \ref{fig:reply_summary}), was decided upon after multiple rounds of internal testing where testers evaluated the messages on whether they contained any of the implications we seek to avoid.

Of course, the steps listed here cannot completely eliminate the 
possibility
of misuse or misinterpretation, and they are not meant as a standalone solution.
Rather, these design choices comprise just one step in our broader response to the ethical challenges of this study, which we continue to discuss in Section \ref{sec:community}.

\subsection{Study Design}
\label{sec:study_design}
Having developed the \convowizard tool as a concrete implementation of the \riskawareness paradigm, we now turn to describe the design of our IRB-approved study in which users tested and gave feedback on \convowizard in real online discussions.

\subsubsection{Community collaboration with \cmv}
\label{sec:community}
As previously mentioned, the need to evaluate our proposed paradigm in a real-world setting raises important ethical challenges, due to the 
danger 
of harm arising from algorithmic flaws or misuse of the \convowizard tool.
While we have taken concrete steps to minimize the possibility of harm
(Section \ref{sec:design_ethics})
such steps can never completely eliminate the 
possibility.

Any harm that does occur might not just be limited to the users of \convowizard---algorithmic flaws or misuse could negatively impact the discussions in which those users partake, and this could have further impacts on the community (subreddit) in which the discussions occur. 
The resulting ethical implication is clear: the potentially affected party, that is the community itself, must be allowed to play an active role in the setup and execution of the study.
This
led
us to develop our study as a \textit{community collaboration}, actively working together with a specific subreddit and giving its members a chance to weigh in.
We specifically chose to collaborate with the subreddit \cmv, a community centered around good-faith debates.
We chose this community for two reasons: it has an established history of research collaborations \cite{jhaver_designing_2017,hidey_analyzing_2017,wei_is_2016,tan_winning_2016},\footnote{These collaborations are publicly promoted on the \cmv community wiki: \url{https://www.reddit.com/r/changemyview/wiki/research}} 
and their overall culture, which prioritizes civility and open-mindedness, is a particularly good fit for our proposed paradigm, which is predicated on the good faith of users.

On Reddit, the term ``moderator'' can be somewhat misleading---volunteer subreddit moderators are not merely responsible for rule enforcement, but rather play a larger social role as \emph{community leaders}, who engage directly with members of the community both formally and informally to build solidarity and construct shared norms \cite{dosono_moderation_2019,seering_reconsidering_2020,gilbert_i_2020} and even serve as their community's representatives to the outside world \cite{seering_metaphors_2020}.\footnote{Concrete examples of this type of work among \cmv moderators include organizing semi-regular town-hall-style feedback threads (\url{https://www.reddit.com/r/changemyview/wiki/metamondays}) and producing a \cmv podcast (\url{https://www.reddit.com/r/changemyview/wiki/podcast}).}
In light of this, our collaboration with \cmv centered around an ongoing dialogue with the \cmv moderators.
We first reached out to them to explain our research and propose a collaboration, and after they collectively agreed to the proposal, we worked together to craft a public announcement explaining the study to the broader \cmv community.
The moderators subsequently posted the announcement as an official pinned post,\footnote{Official pinned posts always appear at the top of the subreddit page and have special styling to visually distinguish them from regular posts.} which through the course of the study served a dual purpose as both a sign-up hub hosting links to join the study, and as a communications hub where \cmv members (whether participating in the study or not) could ask questions, give feedback, or raise any concerns.
As the study proceeded, we maintained our dialogue with the moderators, who acted as intermediaries between us and the \cmv community: they passed along new questions and concerns to us, and we provided them with answers and updates which they could add to the pinned post.

\subsubsection{Experimental design}
\label{sec:experimental_conditions}

\rredit{In order to determine how users might react to \convowizard's interventions, we employ a two-phase user study, consisting of a first phase 
focused on collecting self reports of how participants use \convowizard
, followed by a second phase designed to collect more controlled usage data
for the sake of quantifying the participant-reported effects.
This two-phase design was driven by the aforementioned practical challenge of recruiting regular \cmv users to use \convowizard: as we have described, addressing this practical challenge requires that users perceive \convowizard as providing real value, and a controlled setup can undermine this since \convowizard would not provide any utility to the user within a \control condition.
Having two phases offers a workable compromise, as the uncontrolled first phase allows users to experience \convowizard in full without having to worry about interference from experimental controls, and serves to ease them in to the more complicated (from the user perspective) controlled second phase.}

\rredit{
In Phase 1 of the study, lasting 30 days, participants are asked to install a version of \convowizard that does not implement any experimental controls, thus giving all participants an uninterrupted experience of using the tool.
The focus of this phase is to gather self reports of how participants interact with this new paradigm,
which they provide through an exit survey distributed at the end of the 30-day period (described in more detail in Section \ref{sec:exit_survey}).
}

\rredit{
Phase 2 of the study, lasting 60 days, is designed quantify the participant-reported effects from Phase 1 through a controlled analysis of ConvoWizard usage logs.
To this end, \convowizard in this phase implements} a \emph{within-subjects} \rct design in which we assign \treatment and \control conditions at an interaction level:
when a participant first
hits the ``reply'' button
on a discussion thread within a \cmv post,
\convowizard randomly decides (with probability 0.5) whether or not to show the interventions for 
the user's interactions on that post.
This way we can compare how each participant behaves in the presence (vs. the absence) of the \convowizard intervention.
\rredit{Our choice of a within-subjects design rather than a between-subjects one was again driven by practical considerations:} asking users to install and use a tool that does nothing (as would be the case in the \control setting of a between-subjects study) would be infeasible, whereas the within-subjects design allows every participant to experience \convowizard's functionality at least some of the time.

In total, 47 users finished Phase 1 of the study (including the exit survey) and 14 users finished Phase 2. 
We acknowledge that this results in a self-selected participant pool that is not necessarily representative of the entire \cmv user population, being more likely to attract users that are interested in the issue of incivility.  
\rredit{
Despite this limitation, the resulting data can still be useful as a first step towards characterizing the potential of the \riskawareness paradigm, as we seek to do in the subsequent analysis (Section \ref{sec:findings}).
We return to discuss this limitation---and the steps needed for future work to overcome it---in more detail in Section \ref{sec:discussion}.
}

\subsubsection{Exit survey}
\label{sec:exit_survey}
The exit survey, sent to all Phase 1 participants after the end of the 30-day period, gave participants a chance to report on their experiences with \convowizard and provide their overall impressions of the tool, and serves as an instrument for a qualitative evaluation of the \riskawareness paradigm.
The full text of the survey can be found alongside further details about the execution of the study in Appendix \ref{appendix:exitsurvey}.

The exit survey can be roughly divided into three sections\rredit{, all of which contain a mix of multiple-choice questions and open-ended text responses}.
First, we asked about participants' prior experiences with incivility and moderation, including what effects they think incivility has on discussions, how they personally react to incivility, and how effective moderation has been in their experience.
Next, we asked participants to describe how they tended to respond when \convowizard warned them about risk of incivility, including whether they tended to agree with \convowizard's predictions and whether this subsequently affected their behavior.
Finally, we asked participants to give their general impressions of \convowizard and their willingness to use it in their everyday \cmv participation if it were hypothetically available for general use outside the context of the study.

\subsubsection{Data collection and processing}
\label{sec:data_collection}
As described in Section \ref{sec:prototype_tool}, \convowizard records users' drafting behavior in real time.
This data collection takes place for every interaction regardless of whether the tool is in \treatment mode or \control mode, and the result is a rich record of how users draft their comments both ``naturally'' (in the \control condition) and in the presence of the \convowizard intervention (in the \treatment condition).
In our subsequent analysis we compare the drafting behavior in these two conditions.

To avoid attributing spurious differences to \convowizard, the data must have the following properties:
\begin{itemize}
    \item Each user should contribute an equal number of \treatment and \control interactions.
    This prevents our analysis from uncovering spurious differences arising from individual personality traits of the participants.
    \item The \treatment and \control data should have the same distribution of estimated prior risk.
    This prevents our analysis from uncovering spurious differences arising from how participants react in discussions with different levels of risk.
\end{itemize} 
While with enough participants these properties would follow from the randomization of the experiment design, considering the relatively small number of participants we take an extra step to 
enforce
these properties in our data.
For each logged interaction taking place in the \treatment condition 
(i.e., when \convowizard is active),
we match it with an interaction from the \control condition that was from the same author and had the same level of estimated prior risk (i.e. context risk score).\footnote{The matching algorithm prefers Phase 2 data, but is allowed to draw \treatment data from Phase 1 in the rare case where a \control interaction did not have any valid Phase 2 \treatment match meeting both filter criteria.} 
Any interactions that could not be matched are discarded.
\rredit{This procedure results in a total of 334 pairs (668 total interactions).}

\section{Findings}
\label{sec:findings}
In order to probe the feasibility of our paradigm, we aim to understand whether informing a user that a discussion they participate in is (algorithmically-inferred to be) at risk of derailment will lead them to attempt to mitigate this risk. 
\rredit{Leveraging the mixed methods setup of our study, we address this question by combining qualitative and quantitative insights derived both from exit survey responses and statistical analysis of data collected in the \rct.
In survey responses, participants both report a general willingness to take steps to mitigate risk of derailment, and identify key ways in which \convowizard's algorithmically-provided \riskawareness helps them in that process.
We use these insights to guide an exploratory analysis of how users drafted their comments in the \treatment versus the \control conditions of the \rct.}
More specifically, the rest of this section is organized as follows:
\begin{enumerate}
    \item First, we ask whether, independent of any external assistance, well-intentioned users are already willing to take steps to reduce tension in conversations that they feel are at risk, and if so, what concrete strategies they engage in. 
    Through exit survey responses, we find that most users do self-report that they act proactively to reduce tension, and their responses give further insights into the specific proactive strategies they employ (Section \ref{sec:tension_reduction}).
    \item Next, \rredit{we ask whether users judge risk estimates from an (imperfect) algorithm to be a helpful addition to their own intuitions about risk.}
    Survey responses suggest that the answer is yes: users largely find \convowizard judgments reasonable, \rredit{and point out specific situations in which \convowizard's warnings helped them identify tension that they might not have picked up on otherwise.
    As a further promising sign, users} express a willingness to use the tool as part of their regular Reddit commenting workflow (Section \ref{sec:algorithmic_awareness}).
    \item Finally,
    \rredit{in light of the observation that users are willing to act proactively and find algorithmic input helpful} in deciding when to do so, we seek to \rredit{understand in more detail what these algorithmically-guided proactive steps might concretely look like.
    An initial qualitative picture emerges from freeform survey responses: \convowizard's warnings lead users to reflect further on the tension present in the conversation and how their draft reply might affect it, and to revise their draft reply in ways that might mitigate the risk of escalation.}
    \rredit{Furthermore, a quantitative analysis of participants' comment drafting activity in the \rct reveals effects that, although small, corroborate the aforementioned qualitative findings:
    compared to the \control condition, during the \treatment condition participants tend to spend 
     more time drafting their comments, make revisions that reduce the algorithmically-estimated risk, and shift their language in ways that roughly} correspond to the proactive strategies they reported employing to reduce tension (Section~\ref{sec:editing}).
\end{enumerate}

\subsection{\rredit{Users' Intuitive Perception and Management of Risk}}
\label{sec:tension_reduction}

Responses from the exit survey indicate that in their regular commenting behavior (that is, in the absence of any outside assistance), well-intentioned users already proactively engage in some strategies for reducing tension when they intuitively deem it necessary.
First, every participant reported that they have some level of intuition for when a discussion is at risk of turning uncivil.
\rredit{
Explanations of how this intuition works vary across participants.
Some participants reason about risk in terms of specific word choices:
\begin{quote}
    \participant{7}: Referring to someone as ``you'' tends to signal things may take a turn, as does using generalizing language and absolute terms like ``always'' and ``all''.
\end{quote}
\begin{quote}
    \participant{17}: The easiest way is to analyze the phrasing. Stern, short phrases, completely contradicting the other person's viewpoint might come off as hostile and aggressive, causing a defensive reaction that might turn into an uncivil discussion.
\end{quote}
Meanwhile, other participants look at higher-level concepts such as tone, and especially the sense that an interlocutor is making things personal:
\begin{quote}
    \participant{34}: There is a certain tone or rhetorical posture that people will take prior or during an uncivil reply that forecasts their position. Often times folks that are uncivil also project a greater deal of certainty about their conclusions and will be quicker to disagree or criticize than they are to interrogate the position they disagree with. 
\end{quote}
\begin{quote}
    \participant{33}: The arguments diverge from the topic to trying to guess what the other is supposedly thinking or making assumptions about the person and try to associate them with groups/beliefs etc.
\end{quote}
\begin{quote}
    \participant{18}: [The conversation might be at risk] if the conversation starts getting personal, attacking personal credentials or identity instead of the problem. 
\end{quote}
}
Then, most participants went on to report that this intuition shapes their subsequent behavior: 61.7\% report that they are less likely to join a discussion they suspect to be at risk of incivility, and 76.6\% say that if they do join they will change how they phrase their reply.
The latter finding in particular, that participants change their phrasing in response to perceived risk, is a promising indicator that they are willing to put in effort to reduce, or at least avoid increasing, the tension in the conversation.
We explore this possibility further by examing what specific language changes these participants report.
To focus the scope of this analysis, we specifically consider a set of linguistic phenomena that have been connected to (in)civility and healthy interactions in prior work:
\begin{itemize}
    \item \textbf{Politeness}: Linguists have long theorized that politeness serves as a buffer to soften the perceived force of a message \cite{brown_politeness:_1987,lakoff_logic_1973}, and recent work has empirically validated this \cite{zhang_conversations_2018}.
    \item \textbf{Formality}: Formality has been theorized to play a role in preventing misunderstanding \cite{heylighen_formality_1999}, and in turn misunderstanding has been identified as a potential driver of incivility~\cite{chang_dont_2020}.
    \item \textbf{Objectivity}: In the survey and in this work, we specifically define ``objective'' language as the use of facts and data in constructing a comment, in contrast with the use of personal experiences and emotions. This feature is more specific to our domain of \cmv: we speculate that in the specific context of debates, reliance on fact-driven argumentation may help keep debates on topic and prevent descent into \textit{ad hominems}, which may be connected to incivility \cite{habernal_before_2018}.
    \item \textbf{Question-asking}: Asking more questions might show an interest in engaging with the point of the interlocutor, and has previously been shown to prompt more positive feedback, such as liking and agreement, from interlocutors \cite{huang_it_2017}.
    \item \textbf{Swearing}: Swearing can be used to express aggression, but also to signal group identity or informality \cite{holgate_why_2018}.
    \item \textbf{Comment length}: In the context of debates, higher word count can indicate that the interlocutor is trying to be more explicit in their argument \cite{okeefe_justification_1998,okeefe_standpoint_1997}, which may, like formality, reflect an attempt to avoid misunderstanding.
\end{itemize}
The exit survey asks participants about their use of these strategies in conversations that they intuitively deem to be at risk.
Among them, participants most commonly reported changes in four of them: increased politeness (52.7\% of participants), use of more objective language (66.7\%), asking more questions (50.0\%), and use of more formal language (47.2\%).\footnote{We also offered a free response ``Other'' option so participants could describe strategies that don't fit under any of the listed options. 
A quarter of the
participants took this option, and a 
random
sample of their responses can be found in Appendix \ref{appendix:freeresponse}.}
Overall, these findings show that well-intentioned users desire to avoid escalating \atrisk conversations, and are willing to alter their behavior in order to achieve this goal.
However, this does not imply that they are immune to engaging in uncivil behavior themselves: 68.1\% of participants actually report that they have at some point made a comment that they later regretted because in hindsight it was uncivil.
These regrettable actions may be driven by a number of factors.
For one, sometimes users may be making an inaccurate judgment of risk;%
\rredit{
as \participant{39} puts it:
\begin{quote}
    \participant{39}: It's hard in the moment when reading a divisive comment to objectively recognize where the conversation is going.
\end{quote}
There can also be uncertainty in judging how one's \emph{own} contribution contributes to the risk, as \participant{44} explains:
\begin{quote}
    \participant{44}: I'm not always sure when what I'm going to say will make things better or worse.
\end{quote}
\mredit{
Overall, 78.7\% of participants expressed some degree of uncertainty about their risk intuitions, echoing \participant{39} and \participant{44}'s sentiments.
}%
Furthermore, even if the user \mredit{does make} an accurate judgment, they may \mredit{still} simply get caught up in the heat of the moment \cite{cheng_anyone_2017}.
}

In light of this, we speculate that an additional nudge
that enhances
a user's
awareness of existing tension in the conversation might prevent them from escalating the tension or even from crossing the line into uncivil behavior.
This is a potential opening for algorithmic \riskawareness interventions, and in the following sections we proceed to investigate this possibility.

\subsection{\rredit{Usefulness of Algorithmic Interventions}}
\label{sec:algorithmic_awareness}

\rredit{Our first step in exploring the potential of algorithmic \riskawareness interventions is to check whether users actually find such interventions to be helpful additions to their process of reasoning about risk of incivility.}
\rredit{To this end, we examine participants' exit survey evaluations of their experience with \convowizard, with a particular eye towards how and why they rated its interventions as useful (or not).}\footnote{In the survey, mostly-identical versions of the \convowizard feedback questions were asked separately for the \contextfeed and \replyfeed interventions, to prevent participant confusion. Because the results were broadly similar between the two versions of the questions, to avoid redundancy we will refer in the text to the numbers from the \replyfeed version of the questions (chosen because there are a small handful of questions that were specific to the \replyfeed). 
Full response numbers for both versions of the questions can be found in Appendix \ref{appendix:exitsurvey}.}

We find that participants broadly rated \convowizard's interventions as both useful and intuitively correct: 77.1\% of participants reported that they found the interventions at least somewhat useful, and 68.1\% felt that \convowizard's estimates of risk were as good as or better than their own intuition.
Furthermore, responses suggest that many participants see acting upon \convowizard's warnings as being to their benefit: over half of the participants felt that \convowizard's warnings stopped them from engaging in fights with other interlocutors during the experimental period (54.3\%), and even prevented them from posting a comment they would have later regretted (54.3\%).\footnote{In interpreting these percentages, one should consider that not all participants are expected to be in a situation where they are about to enter a fight or post a regrettable comment during the experimental period.
}

\rredit{
To put these numbers in more context, we examine participants' open-ended responses, which shed light on exactly \emph{how} \convowizard helped them.
In these responses, participants identify a number of ways in which \convowizard made them more aware of tension in conversations and in their draft replies.
Some participants felt that \convowizard performed \emph{better} than their own intuition at detecting risk, in that it picked up on cases of tension that they would have missed. \participant{18} explains:
\begin{quote}
\participant{18}: I feel like I don't pay attention to specific triggers programmed into the wizard. Even if my message isn't confrontational the way I say it might have an unintended psychological impact I wouldn't have recognized. 
\end{quote}
For other participants, even if \convowizard was not necessarily better than their own intuition, it served as a second opinion providing clarity in cases where their intuition left them uncertain, as \participant{13} found:
\begin{quote}
\participant{13}: In situations in which I would need more context to see where the discussion is going, ConvoWizard's answer is `yes' or `no' while mine is `I don't know yet', and it's usually right still.
\end{quote}
Finally, for some participants \convowizard played a somewhat more modest but still impactful role: it served as a prompt to think about tension in cases where they wouldn't have been thinking about it. \participant{15} elaborates:
\begin{quote}
\participant{15}: I don't often care about increasing tension. My objective is generally the discussion, not whether I sound polite or not. ConvoWizard sort of reminds me that I should use maybe different language.
\end{quote}
Thus, while individual participants might differ in exactly how they benefited from \convowizard's interventions, on the whole we find that \convowizard fills various gaps in their reasoning about tension and thus serves to increase their overall awareness of risk.
}

Another important factor in judging 
\convowizard's usefulness is participants' willingness to continue using it if it were made widely available.
Here, we find that 83.0\% of participants expressed at least some interest in adopting \convowizard as part of their usual \cmv workflow, if it were publicly deployed.
Perhaps more importantly, 63.8\% of participants felt that if \convowizard were to be broadly adopted by the \cmv community, the net effect would be an \emph{improvement} in discussion quality.

\rredit{
Taken together, these results are a promising initial sign that algorithmic \riskawareness interventions can be a valuable tool to help users identify tense conversations.
That said, it is just as important to note that as an early prototype, \convowizard is far from perfect, and participants also identified specific shortcomings that prevented it from being as useful as it could have been.
Most notable among these is the issue of \emph{false positives}: when participants were asked about reasons they might sometimes disagree with \convowizard's judgments, false positives were a more commonly cited concern than false negatives, with 61.7\% reporting that the former was a common issue they encountered, and only 34.1\% reporting the latter.
False positives can detract from the overall helpfulness of the tool since too many unwarranted warnings can make the tool seem annoying, as \participant{2} explains:
\begin{quote}
    \participant{2}: The ``false positive'' rate was much higher than the ``false negative'' rate [\dots] This was helpful in detecting some things that ought to be rephrased, but slightly annoying at times after several re-edits of the intended comment.
\end{quote} 
In the extreme, it could also lead to a boy-who-cried-wolf situation, in which users end up dismissing the tool as just always reporting tension regardless of what is actually happening in the conversation, as \participant{37} succinctly puts it:
\begin{quote}
    \participant{37}: It seemed to say everything was in danger of tension
\end{quote}
These observations mark an important direction for future work.
While ideally tools like \convowizard would benefit from improved algorithms that make fewer false positive errors, in light of the fact that the algorithm will never be perfect there is a potential design implication here: future work could look into ways to better trade off precision and recall, or even offer users intuitive ways to adjust this tradeoff to their own preferences.
}

\rredit{
Another important drawback that participants identified was lack of transparency: 48.9\% of participants marked ``more transparency'' as one of the most important improvements they would want to see in a future iteration of \convowizard.
The lack of transparency limits \convowizard's helpfulness in two key ways.
First, as \participant{11} explains, it can leave users knowing that a conversation is at risk but not knowing what to do about it:
\begin{quote}
    \participant{11}: I think it needs to get better at walking through why it thinks a thread is hostile and why your reply is. It was often left to me to entirely rethink a statement which seemed to say it was better without explaining why that change helped.
\end{quote}
Second, similar to the issue that was raised in the discussion of false positives, seeing the algorithm make apparent mistakes with no explanation as to why can eventually lead the user to tune out the tool's feedback, a situation that \participant{13} identifies:
\begin{quote}
    \participant{13}: Since the reply summary feature flipflopped regularly, I ended up not paying a lot of attention to it. So probably also in cases in which it would have been helpful.
\end{quote}
Future 
implementations
 should therefore 
seek to integrate
recent developments in explaining algorithmic decisions (as seen with toxicity detection, for instance, in the RECAST system \cite{wright_recast_2021}) to build algorithmic risk awareness interventions that are more explainable and hence, perhaps, more directly actionable.
}

\rredit{
Overall, while there is clearly more work needed to help algorithmic \riskawareness tools meet their full potential, as a preliminary step the results of our study serve to establish that such tools are at least feasible as a means of increasing users' awareness of risk in conversations.
Having established this, we now turn to investigate the implications of this increased awareness; that is, what concrete steps users might take to mitigate risk when it is brought to their attention.
}

\subsection{\rredit{How Users Engage With Algorithmic Interventions}}
\label{sec:editing}

Our 
exploration of 
\rredit{how users engage with the enhanced \riskawareness provided by algorithmic interventions}
is guided by prior work on user-facing interventions aimed at promoting prosocial behavior.
Specifically, we focus our attention on two types of concrete%
\rredit{
proactive steps users might take: spending extra time to consider and react to \convowizard's warnings while writing their comment \cite{kriplean_is_2012}, and making (token-level) adjustments to their language use \cite{seering_designing_2019}.
In addition to looking for
self-reports
of such reactions in the survey responses, we also seek to support any self-reported findings with
evidence from the experimental data, by running comparative \control-versus-\treatment analyses at the \emph{interaction} level (that is, on the 668 paired interactions described in Section \ref{sec:data_collection}).}

\rredit{
We note, however, that 
the design of the study imposes several limitations on the comparative analysis: the small sample size restricts us
to the use of coarse-grained, simplified metrics and 
necessarily leads to
 low-powered results,
and the within-subjects setup prevents us from inferring broader behavioral changes beyond how users immediately engage with system interventions.
As such, the results should best be understood as highlighting potentially interesting trends in order to guide subsequent work, rather than as being exhaustively conclusive in and of themselves.
}

\subsubsection{\rredit{Deeper reflection and revision}}
\label{sec:drafting_effects}

\rredit{One basic way users might engage with algorithmic warnings of risk would be to spend more time to consider the tension being pointed out by the algorithm and think about how to reword their comment accordingly.}
This kind of effect was previously shown in \citeauthor{kriplean_is_2012}'s work on the ``Reflect'' intervention, where users reported taking the time to more deeply consider the comment they were replying to, which the authors speculated would ``act to counteract our tendency towards knee-jerk reactions'' \cite{kriplean_is_2012}---precisely the kind of impact we sought to achieve with \convowizard.

\rredit{
In open-ended responses, several participants indeed report engaging in such reflection and revision.
For instance, \participant{26} points out how seeing a warning from \convowizard might prompt them to review the conversational context more deeply than they would otherwise:
\begin{quote}
    \participant{26}: I don't always read the entire chain of parent comments so the wizard indicating concern lead me to go back and read the entire chain. 
\end{quote}
\participant{9} notes that even though they were aware the algorithm is imperfect, it was good enough to prompt reflection on their own in-progress draft:
\begin{quote}
    \participant{9}: I'm sure its not perfect, but in my case it made me rethink what I type.
\end{quote}
Finally, \participant{2} explicitly mentions spending extra time rewording their comments:
\begin{quote}
    \participant{2}: I spent a lot of time rephrasing.  Often there were phrases that in other contexts could signal increasing tension, but would not in the context I typed.  
\end{quote}
}

\begin{table}
    \flushleft
    \begin{subtable}[b]{0.47\textwidth}
        \begin{tabular}{rr|c}
            & & Drafting time \\
            & & (seconds) \\
            \hline
            \multirow{2}*{\Atrisk} & \control & 174.2 \\
            & \treatment & \textbf{189.5} \\
            \hline
            \multirow{2}*{\Notatrisk} & \control & 124.3 \\
            & \treatment & 133.4
        \end{tabular}
        \caption{}
        \label{tab:draft_time}
    \end{subtable}
    \begin{subtable}[b]{0.45\textwidth}
        \centering
        \begin{tabular}{rr|c}
            & & Correlation between \\
            & & adjusted timestamp \\
            & & and risk score \\
            \hline
            \multirow{2}*{\Atrisk} & \control & $\ 0.05^{**}$ \\
            & \treatment & $-0.06^{***}$ \\
            \hline
            \multirow{2}*{\Notatrisk} & \control & $-0.13^{***}$ \\
            & \treatment & $-0.06^{**}$
        \end{tabular}
        \caption{}
        \label{tab:score_correlation}
    \end{subtable}
    \caption{\control-versus-\treatment comparisons of two high-level measures of drafting behavior: (a) Average time spent per 
    interaction,
     in seconds. \textbf{Bolded} \treatment values are significantly ($p < 0.05$, Mann-Whitney test) different from their \control counterparts. (b) Correlations between adjusted timestamp (time in seconds since the start of the interaction) and risk score (as determined by CRAFT). Correlations are measured as Spearman's R and stars indicate significance levels ($^{**}p < 0.01, ^{***}p < 0.001$).%
     }
    \label{tab:engagement}
\end{table}

\rredit{
Since reflection is an inherently subjective process we cannot quantify it directly. We can however check for the existence of the time effect that we would expect to accompany increased reflection (and which \participant{2} explicitly calls out).
To this end,} we compute the mean time spent per logged interaction, stratified both by experimental condition and by
whether the interaction was judged to be \emph{\atrisk} (i.e., there was enough 
algorithmically-inferred
 tension that, for \treatment interactions, a warning was displayed, and for \control interactions, a warning would have been displayed had \convowizard been active).
If the hypothesized engagement effect exists, we expect that there should be an increase in average time per interaction in the \treatment condition---but importantly, because the hypothesized effect is a response to \convowizard's warnings, we expect this difference to exist only in \atrisk interactions 
(since that is the only scenario in which \convowizard would display an intervention in the \treatment condition and not display one in the \control condition).

We find this exact effect: in \atrisk interactions, there is a significant ($p < 0.05$ via Mann-Whitney test) increase in the mean amount of time spent per interaction in the \treatment condition, and no such change in \notatrisk interactions (Table \ref{tab:draft_time}).
We further note that this increase cannot simply be explained by differences in the length of the comments; in fact, the average number of words per comment does not differ significantly between the two conditions ($p = 0.21$ via Mann-Whitney test).  
\rredit{
While 
we reiterate that
this analysis cannot directly measure how much participants actually reflect on their drafts, it 
at least 
offers some quantitative corroboration of their self-reports.
}

\rredit{
As a next step, we want to know how this engagement with the tool might translate into concrete changes to the drafting and revision process. 
To investigate this, we again start from the open-ended responses: 
some participants report that they use \convowizard's risk intensity feature (i.e., the changing colors indicating levels of estimated risk) as a guide, attempting to revise their comment in a way that produces a less intense color (i.e., lower risk score):
\begin{quote}
    \participant{35}: I kept rewording my reply until it stopped showing up orange.
\end{quote}
\begin{quote}
    \participant{9}: Often times if the color changed I would reread what I was saying and see if the response maybe came off the wrong way. Helping me then to reword it.
\end{quote}
In the 
\rct data, if users are actively attempting to reduce the 
degree of tension displayed by ConvoWizard,
 as suggested by \participant{35} and \participant{9}, then in the \treatment condition we would expect to see a gradual decrease in the risk score as a comment gets drafted; that is, we expect an inverse correlation between the risk scores of the intermediate snapshots of a draft and their associated timestamps.} 

We indeed find (Table \ref{tab:score_correlation}) that in \atrisk interactions in the \treatment condition (i.e., interactions where a warning was displayed), there is a negative correlation between timestamp and risk score.\footnote{We normalize by the timestamp at which the interaction started (i.e., adjusted timestamp = timestamp $-$ timestamp of first snapshot in this interaction, such that the adjusted timestamp of the first snapshot is always 0).}
Notably, the corresponding correlation for the \textit{\control} condition is actually \textit{positive}.
In other words, in \atrisk situations the natural tendency is for risk score to \emph{increase} over time as a draft is written, and the introduction of the \convowizard intervention actually manages to reverse this natural trend.
\rredit{
While the correlations themselves are relatively small in magnitude, it should be noted that a rank-order correlation test is a very coarse metric of the phenomenon being investigated here, since an algorithmic warning and subsequent risk-score-decreasing edit could occur at any point---or even \emph{multiple} points---in the drafting process, so the true relationship may not be monotonic over the entire duration of the interaction.
In this sense,
it is promising that even such a coarse metric can reveal a significant trend,
and this
suggests the potential for more sophisticated analyses. 
For example, a larger study could allow for a more precise analysis considering the exact moment of each warning and the subsequent edits it triggers.
}

\begin{table}
    \begin{tabular}{rr|c|c|c}
        & & Formality rate & CDI (mean) & Question rate \\
        \hline
        \multirow{2}*{\Atrisk} & \control & 81.8\% & 0.06 & 15.3\% \\
        & \treatment & \textbf{87.1\%}& \textit{0.09} & \textbf{20.4\%} \\
        \hline
        \multirow{2}*{\Notatrisk} & \control & 92.8\% & 0.12 & 11.6\% \\
        & \treatment & 92.1\% & 0.11 & 14.3\% \\
    \end{tabular}
    \caption{
        \control-versus-\treatment comparisons of three linguistic strategies: formality (measured using the discretized F-factor), the categorical-dynamic index (CDI, used as a rough proxy for objectivity) and the rate of question-asking. \textbf{Bolded} \treatment values are significantly ($p < 0.05$) different from their \control counterparts, while \textit{italicized} results indicate an almost-significant trend ($p = 0.07$). Significance is tested using Mann-Whitney for comparison of means, and 
        \rredit{Fisher's exact test} for comparison of rates.}
    \label{tab:language_results}
\end{table}

\subsubsection{Effects on linguistic strategies}
\rredit{
Once a user has reflected on the tension identified by an algorithmic intervention, and revised their comment accordingly, does this end up being 
echoed
 in the language of the reply they end up posting?
Broadly speaking, participants self-report that this is the case, with 71.4\% reporting that \convowizard warnings affected the language they used in their replies---but what do these changes specifically consist of?
}

\rredit{
In exploring this question, we must keep in mind that users are somewhat constrained in the extent to which they can alter their language, since ultimately the goal of the conversation is to have a debate and so users cannot make drastic changes that would alter the semantic content of their comment.
As such, to the extent that linguistic change occurs, it is aimed at controlling the \emph{tone} that gets conveyed while preserving semantic meaning, as \participant{9} and \participant{18} explain:
\begin{quote}
    \participant{9}: I thought of better words I could use maybe words that don't sound like I may be trying to provoke a uncivil response.
\end{quote}
\begin{quote}
    \participant{18}: I tended to avoid certain key words that I felt the program picked up on whether or not I was being confrontational. The word ``you'' or any words with negative connotations could be altered without changing the meat of my messages. 
\end{quote}
More 
concretely, participants reported that ConvoWizard warnings led to increases along the same four linguistic strategies that they had previously identified as their reactions to 
intuitively perceived risk:
 politeness (68.0\% of participants who reported any linguistic changes), formality (48.0\%), objectivity (44.0\%), and question-asking (32.0\%).
}

\rredit{These results inform our subsequent comparative analysis of linguistic effects
in the \rct.
As with our earlier analyses,} given the limited size of the controlled data, we are necessarily limited in the complexity of the linguistic phenomena we can capture in our analysis. 
To this end, we adopt a similar strategy to that used by \citeauthor{seering_designing_2019} in their work on interventions for encouraging prosocial behavior: comparing basic \emph{summary variables} that can be computed as simple functions of tokens and parts-of-speech \cite{seering_designing_2019}.
Our specific choice of summary variables is inspired by\rredit{---but not exhaustive of}\footnote{Notably, we do not consider politeness, since to the best of our knowledge no trained model exists for \cmv comments and additional labeled data would be needed to train such models (existing politeness models are trained on \emph{requests} extracted from Wikipedia Talk Pages and StackExchange comments \cite{danescu-niculescu-mizil_computational_2013}).}\rredit{---}the strategies 
that 
users self-reported 
employing 
in order to reduce tension:
\begin{itemize}
    \item \textbf{F-factor}: This is a simple measure of \emph{formality} introduced in \cite{heylighen_formality_1999}. It is computed as:
    \begin{equation*}
    \begin{aligned}
    F ={} & (\text{freq(nouns)} + \text{freq(adjectives)} + \text{freq(prepositions)} + \text{freq(articles)} \\
    & - \text{freq(pronouns)} - \text{freq(verbs)} - \text{freq(adverbs)} - \text{freq(interjections)} + 1) / 2
    \end{aligned}
    \end{equation*}
    Where freq() measures the frequency of a given word category in a body of text; that is, a count of words of that type normalized by the total number of words in the text.
    Because the short length of Reddit comments makes the F-factor somewhat noisy, we  
    use a discretized version of the score, adopting an empirical threshold of $0.44$ that was inferred by the authors of the F-factor based on an analysis of labeled corpora. F-factor is thus discretized as simply ``informal'' ($F \le 0.44$) or ``formal'' ($F > 0.44$).
    These discretized scores are compared as ``formality rates''; that is, the percentage of all comments that get scored as ``formal'' within a given set of comments.
    \item \textbf{Categorical-Dynamic Index (CDI)}: This is a 
    score derived from function word counts, with
     higher values indicating a more analytic and cognitively complex writing style, and lower values indicating more reliance on storytelling and personal narratives \cite{pennebaker_when_2014}.
    We use this score as it roughly corresponds to our definition of the \textit{objective-subjective} distinction. 
      It is computed as:
    \begin{equation*}
    \begin{aligned}
    \text{CDI} = {} & 0.3 + \text{freq(articles)} + \text{freq(prepositions)} - \text{freq(personal pronouns)} \\
    & - \text{freq(impersonal pronouns)} - \text{freq(aux. verbs)} - \text{freq(conjunctions)} \\
    & - \text{freq(adverbs) - \text{freq(negations)}}
    \end{aligned}
    \end{equation*}
    We note that the CDI is one of the metrics used for quantifying the effects of prosocial interventions in \cite{seering_designing_2019}.
    \item \textbf{Question Rate}: This simply computes what fraction of all sentences within a collection of text are \textit{questions}. While in theory there can be some nuance in what makes a sentence a question, prior computational work on questions found that the simple heuristic of checking for a question mark works remarkably well \cite{zhang_asking_2017}, and so we adopt this heuristic.
\end{itemize}

Table \ref{tab:language_results} shows the results of comparing each variable in the \treatment and \control, stratified by whether the interaction was \atrisk or not.
We find a number of notable differences in the comparisons.
Compared to users in \control, users in \treatment ask more questions and are more likely to write comments that are judged as ``formal'' (according to the discretized F-factor);\footnote{We note that the raw F-factor comparison is not significant, potentially due to the noisiness described before.} both these differences are significant at $p < 0.05$.
Furthermore, like the drafting effects, these effects are only found in \atrisk interactions,\footnote{The apparent increase in questioning in \notatrisk interactions as well is not statistically significant.} suggesting that,
as expected,
they are specific reactions to warnings.
Finally, we find a similar trend (bordering on significance) in the CDI scores, with users in \treatment writing comments with a higher CDI; this is again specific to \atrisk interactions.

\rredit{
As was the case with earlier analyses, these differences, while significant, are relatively small in magnitude.
To some degree this is expected, since as explained earlier the goal-oriented nature of \cmv discussions constrains the extent to which users can alter their language.
That said, the simplistic nature of the language features being measured here may also play a role in the effect sizes we are observing.
In particular, while for the sake of accomodating our limited data we specifically chose lexically-derived features, participant responses previously quoted in Section \ref{sec:tension_reduction} suggest that the most informative linguistic signals of tension and lack thereof, such as tone and making things personal, may not be so easily captured at the lexical level alone.
As such, a future larger-scale study could aim to collect enough data to enable analysis using more sophisticated NLP approaches, which could better capture such high-level phenomena---and in the meantime our preliminary results here suggest that linguistic effects are, in fact, a promising target for such continued exploration.
}

\rredit{
Taken together, these combined qualitative and quantitative findings 
support %
 a potential mechanism through which the \riskawareness paradigm can 
 contribute to
  more civil online discussions: warnings can lead users to  reflect more deeply about the impact their replies have on their conversations and to revise the language of their draft in a way that reduces the risk of derailment.  
These findings suggest concrete directions for both the design and evaluation of future implementations of the paradigm.
From a design perspective, future implementations could explore additional functionality to support the reflection and revision process; for example, using human-readable explanations (as discussed in Section \ref{sec:algorithmic_awareness}) to guide revisions in a more directly actionable way.
From an evaluation perspective, larger-scale studies are needed to measure the reflection and revision effects in more nuanced and robust ways, including taking a more 
fine-grained look at the drafting process
 to capture immediate responses 
 to warnings,
using more advanced NLP techniques to capture more abstract changes in language,
and running a between-subjects assignment to enable analysis of broader behavioral changes.
These future steps would build upon the groundwork established by our current preliminary study, and thereby bring the risk awareness paradigm closer to its full potential.
}

\section{Discussion}
\label{sec:discussion}
This work starts from the viewpoint that the solution to incivility in online discussions should come, in part, from the participants in these discussions.
They can---and, as they indicate in our exit survey, do---use their conversational skills to proactively reduce tension when they are aware that the discussions they engage in may be at risk of derailing into uncivil behavior.
However, they sometimes also miss the opportunity to react and use these prosocial skills, in which case they may end up escalating the tension or even reply with an uncivil comment they later regret posting. 
Starting from this premise, we propose a new proactive paradigm which seeks to prompt participants to employ their prosocial conversational skills by enhancing their awareness about the risks of the discussions they engage in.
To demonstrate the potential this paradigm has in a real world setting, we developed an algorithmic tool that can inform a user about existing tension in their conversation and in their reply draft in real time, and conducted a user study in a popular debate community.
The results show that users are indeed responsive to the additional \riskawareness provided by our tool: the tool's warnings prompt participants to spend more time (re)considering 
their language,
 and activate conversational skills that they normally employ to reduce tension in conversations.

Unlike solutions that rely solely on moderators, the \riskawareness paradigm is decentralized and thus can more easily scale with the number of users on the platform.
As such, tools based on this paradigm could be a valuable addition to the broader arsenal of moderation strategies employed by online communities.
However, fully deploying such tools at scale requires first carefully understanding the impacts they might have on users and the community as a whole.
\rredit{Our present work takes an important first step towards this understanding, using a small-scale study to establish the necessary groundwork for subsequent larger scale follow-ups and identify specific directions that such future work should pursue more deeply,
as we discuss below.}

\xhdr{Model error and ethical considerations} Any tools interfering in online discourse through algorithmic means should be subject to ethical scrutiny.  
Unlike paradigms that seek to outright automate the moderation process, our approach aims to merely provide information to the users, and does not trigger harsh actions such as content removal or user banning. 
Nevertheless, tools like \convowizard still have an inherent potential for negative consequences due to their reliance on imperfect algorithms---giving users erroneous information about the risk level in their conversations could cause harm, especially if these errors arise from model bias against marginalized groups.

It must further be noted that even in the absence of model error, there are still ethical concerns at a more conceptual level.
While the \riskawareness paradigm aims to improve the civility of online discourse, ``civility'' is ill-defined and often varies by community \cite{chandrasekharan_internets_2018}, and there can be a fine line between incivility and mere disagreement \cite{arazy_stay_2013}.
As such, the \riskawareness paradigm---like other moderation strategies---may risk creating a chilling effect on speech that disincentivizes users from expressing disagreement at all \cite{gillespie_expanding_2020} or ``tone policing'' the type of disagreement that does end up happening, restricting free expression in a way that might systematically silence certain social groups \cite{gorwa_algorithmic_2020}.
These concerns are exacerbated by the observation that the lines between incivility and disagreement are especially likely to get blurred in debates over contentious or controversial topics \cite{crawford_what_2016}, which are exactly the cases where it is particularly important to make sure that already-marginalized voices are not further silenced.

We have been cognizant of these potential harms in designing our study, and the need to account for them ended up shaping key parts of our study design, such as purposely avoiding prescriptive and blame-assigning language (Section \ref{sec:design_ethics}) and running our study as a collaborative effort with community input (Section \ref{sec:community}).
However, further work is needed both to more rigorously characterize the potential harms that can arise from erroneous risk level estimates, and to explore further ways of mitigating these harms.
In particular, future work should look into ways to make algorithmic \riskawareness interventions more \emph{transparent} and \emph{explainable}, which could shed light on algorithmic biases and help users make more informed decisions about each individual intervention \cite{wright_recast_2021}.

\xhdr{\rredit{Well-intentioned users}}
As we have previously described, our proposed \riskawareness paradigm is designed to be used by well-intentioned users\rredit{---that is, those ``ordinary'' users who seek to engage with and contribute to their community in good faith, as opposed to deliberately seeking conflict, and who comprise the majority of users within many communities including \cmv.}
While our exit survey results suggest
that participants in our study meet this description,
\rredit{we must acknowledge that self-selection effects likely resulted in a participant pool that is not necessarily representative of well-intentioned users in general; specifically, users who are willing to volunteer for a study on civility may do so because they are unusually thoughtful about civility compared to the average well-intentioned user.
In order to move beyond the proof-of-concept stage, future work would need to look into ethically viable ways to scale up testing and evaluate the effectiveness of tools like ConvoWizard in the hands of a more general pool of users who, while still well-intentioned in the sense of not being bad actors, may be less deliberately reflective of tension compared to the participants in our small, self-selecting pool.}

\rredit{
Beyond study limitations, a separate concern regarding our \riskawareness paradigm's reliance on well-intentioned users might arise when thinking about possible future real-world deployment.
While we have argued that most users are well-intentioned, bad actors exist in any community and can misuse publicly available moderation tools towards malicious ends \cite{jhaver_did_2019}.
A public deployment of a tool like ConvoWizard would likewise be vulnerable to misuse; for example, as described in Section \ref{sec:design_ethics}, a bad faith troll could deliberately attempt to craft a message that triggers a warning.
}

One initial response to this concern is to point out that a similar premise of good faith underlies a number of user-facing moderation tools that already see widespread, large scale use---for example, both community voting \cite{lampe_slashdot_2004,mamykina_design_2011} and flagging/reporting systems \cite{crawford_what_2016} only work to counteract incivility if they are used by users who actually desire civility, and are theoretically vulnerable to abuse by bad-faith users \cite{richterich_karma_2014}.
This has not stopped such systems from becoming a common part of platforms' moderation toolboxes---they are simply not the \emph{only} tools in those toolboxes~\cite{seering_reconsidering_2020}.
We similarly envision tools like \convowizard being integrated into a broader moderation ecosystem, 
\rredit{which could provide ways of establishing checks and balances against misuse.} %
In particular we expect that moderators---who are best positioned to determine what ``well-intentioned'' means in the context of their community---could retain a degree of control over the deployment of these tools\rredit{, in a similar way to how we controlled access to the ConvoWizard prototype to minimize the potential of misuse within the context of the study.}
For instance, moderators may decide whether \riskawareness tools are a good fit for their community at all (as we will further discuss below), or even take a finer-grained approach and set limits on who can access the tool, perhaps using hand-written rules and heuristics
\rredit{(e.g., a minimum activity filter similar to the one we implemented in our study recruitment)}
in a system like Reddit AutoModerator \cite{jhaver_human-machine_2019}.
In light of this, a natural next step for future work might be to conduct a study with moderators to get insights on how they might manage the deployment of tools like \convowizard, and what concrete features would need to be implemented to meet their use case.

\xhdr{Downstream effects} 
This work has characterized the effect of \convowizard's warnings on how its users draft their replies.
But a reply does not exist in a vacuum---it is part of a larger discussion, and so a change in the language of one reply might have further downstream effects on subsequent replies and on the outcome of the discussion.
Future work should investigate such downstream effects, with a particular eye on whether the prosocial changes triggered by a \convowizard warning (Section \ref{sec:drafting_effects}) might further translate to more civil behavior of other interlocutors \cite{bao_conversations_2021},
or whether they strengthen or weaken the persuasive effectiveness of the argument \cite{tan_winning_2016}.
An even larger scale study could additionally examine community-level effects, looking for empirical support of participants' self-reported belief that wide adoption of a tool like \convowizard would improve the quality of discourse in the community (Section \ref{sec:algorithmic_awareness}).

\xhdr{Further domains and use cases}
Our study has focused on one community, \cmv, which was specifically selected because 
it aims to host good faith debates \cite{tan_winning_2016}.
This naturally leads to questions about how well a tool like \convowizard would generalize to other communities.
\rredit{Given the aforementioned 
targeting of well-intentioned users,
 it is fair to acknowledge that our paradigm has little value in communities where such 
  users are sparse.}
Nevertheless, we believe that there are other communities with similar values to \cmv where the \riskawareness paradigm could be very impactful.
In particular,
\emph{goal-oriented} communities, including Q\&A communities like StackOverflow and Quora \cite{mamykina_design_2011} as well as work-coordination settings like Wikipedia Talk Pages \cite{wulczyn_ex_2017,kittur_harnessing_2008}, have an added incentive to keep discussions civil since incivility can distract from their broader non-conversational goals~\cite{arazy_stay_2013}.
Future work could conduct follow-up studies on such platforms
to better understand how the specific needs of these communities might differ from those of 
debate-centric
 communities like \cmv, and what implications these community-specific needs might have on the implementation and effectiveness of the \riskawareness paradigm.

\section*{Acknowledgements}

We would like to thank Michael Bernstein, Lillian Lee, Karen Levy, Cecelia Madsen, the 2021-2022 cohort of fellows at the Center for Advanced Study in the Behavioral Sciences at Stanford, and all the reviewers for the enlightening discussions and helpful suggestions.
We additionally recognize everyone who helped with the implementation of \convowizard, particularly Lucas Van Bramer and Oscar So for their contributions to the codebase, Todd Cullen for his help in setting up the backend server configuration, and Caleb Chiam, Liye Fu, Khonzoda Umarova, and Justine Zhang for their extensive testing and generous feedback.
Finally, we are grateful to all the \cmv users who participated in our study, and to the \cmv moderators for their guidance and for serving as a point of contact to the broader \cmv community.
This research was supported in part by an NSF CAREER award IIS-1750615 and by an NSF Grant IIS-1910147; Jonathan P. Chang was supported in part by a fellowship with the Cornell Center for Social Sciences and Cristian Danescu-Niculescu-Mizil was supported in part by fellowships with the Cornell Center for Social Sciences and with the Center for Advanced Study in the Behavioral Sciences at Stanford.

\bibliographystyle{ACM-Reference-Format}
\bibliography{cscw2022_userstudy_jpc_autoupdate}

\appendix

\section{User Study and Exit Survey Details}
\label{appendix:exitsurvey}

\subsection{Participant Recruitment}
Participants for the study were recruited through two channels.
First, the pinned announcement on \cmv contained links for interested users to sign up for the study.
Second, we direct messaged active members of \cmv.
Regardless of recruitment channel, all potential participants underwent a basic check of prior activity on \cmv to filter out possible sockpuppet or brigader accounts, and also had to fill out a basic eligibility check to make sure that their typical \cmv usage was compatible with \convowizard's technical limitations.\footnote{In particular, \convowizard's DOM-manipulation code was specifically engineered around the HTML structure of Reddit's classic desktop interface (``Old Reddit'') and only works there, so users who primarily use other platforms (e.g., mobile) to access \cmv would be ineligible.}
As an incentive for participation, \$20 Amazon gift cards were offered to all participants who completed Phase 1 of the study, including filling out the exit survey.
\mredit{Across all participants who completed Phase 1, the mean \emph{community age} (i.e., how long they had been active on \cmv by the time of the study) was 3 years; the minimum was 3 months and the maximum was 8 years.}

After Phase 1 was completed, we direct messaged all participants who had indicated in the exit survey that they would be interested in a follow-up study, inviting them to participate in Phase 2.
For Phase 2, participants were given the option of participating for either a 30-day period (for which a \$30 gift card incentive was offered) or a 60-day period (for which a \$70 gift card incentive was offered).
All participants who accepted the invitation to join Phase 2 chose the 60-day option. 

\subsection{Exit Survey Implementation}
The exit survey was implemented as a Qualtrics form, mostly consisting of multiple-choice questions with some optional free-response areas for participants to elaborate on their answers.
The \convowizard tool automatically served the survey link to participants at the end of the 30-day period and participants could fill it out at any time after that, though we did send reminders via Reddit direct message.

\subsection{Exit Survey Full Text and Raw Response Counts}

\textit{Total Responses: 47}

\noindent\underline{ConvoWizard Exit Survey}

Thank you for your participation in the ConvoWizard study! As the final step in the study, we will now ask you a series of questions regarding your experience with ConvoWizard. The survey consists of a mix of multiple choice and free response questions. For free response questions, please provide as much information as you can. Your insights are extremely valuable in helping us with our research and, ultimately, with improving ConvoWizard.

After you submit this survey, we will follow up with your reward for participation (a \$20 Amazon gift card) via DM to the Reddit account you used to sign up for this study.

\begin{enumerate}[start=1,label={\bfseries Q\arabic*:}]
    \item To begin, please enter your Reddit username. \textit{[Free response]}
\end{enumerate}

\noindent\underline{Part 2: Experiences with incivility on r/changemyview}

The following questions will ask about your experiences with uncivil behavior on r/changemyview. For the purposes of this survey, ``uncivil behavior'' can be understood as comments that you judge to be violations of r/changemyview's Rule 2,* regardless of whether they ended up getting removed by moderators.\\
*Rule 2 says ``Don't be rude or hostile to other users. Your comment will be removed even if the rest of it is solid. `They started it' is not an excuse. You should report, not retaliate."

\begin{enumerate}[start=2,label={\bfseries Q\arabic*:}]
    \item How big of a problem do you think incivility is on r/changemyview?
    \begin{itemize}
        \item It is almost nonexistent.: \textit{3}
        \item It is only a minor problem.: \textit{10}
        \item It is noticeable but not too big a problem.: \textit{26}
        \item It is a pretty big problem.: \textit{5}
        \item It is one of the biggest problems on the subreddit.: \textit{3}
    \end{itemize}
    \item In your experience, what \textit{most commonly} happens to uncivil comments on r/changemyview? 
    \begin{itemize}
        \item I don't know (I have never seen any uncivil comments).: \textit{2}
        \item They are removed by moderators.: \textit{32}
        \item They are removed by the author.: \textit{1}
        \item Nothing happens (the comment stays up).: \textit{12}
    \end{itemize}
    \item In your experience, how quickly do r/changemyview moderators take action on uncivil comments?
    \begin{itemize}
        \item I have never seen moderators take action on uncivil comments.: \textit{3}
        \item They act almost immediately after the comment is posted.: \textit{4}
        \item They act within a few hours after the comment is posted.: \textit{25}
        \item They act within the day the comment is posted (but take more than a few hours).: \textit{13}
        \item They take more than a day to act.: \textit{2}
    \end{itemize}
    \item In your experience, what \textit{most commonly} happens to discussions on r/changemyview after an uncivil comment gets posted and is not immediately removed?
    \begin{itemize}
        \item I don't know (I have never seen any uncivil comments, or every uncivil comment I've seen was immediately removed).: \textit{2}
        \item The situation escalates and more uncivil replies are posted.: \textit{22}
        \item The situation recovers and becomes civil again.: \textit{5}
        \item The discussion dies and no further replies are posted.: \textit{18}
    \end{itemize}
\end{enumerate}
\begin{adjustwidth}{0.75in}{}
\textit{Show the following question(s) if ``The situation escalates and more uncivil replies are posted'' was selected in \textbf{Q5} (22 participants):}
\begin{enumerate}[start=6,label={\bfseries Q\arabic*:}]
    \item In discussions that you've seen escalate after an uncivil comment was posted and not immediately removed, what \textit{most commonly} happens if the comment is eventually removed?
    \begin{itemize}
        \item I don't know (I have never seen an uncivil comment get removed).: \textit{1}
        \item The removal helps the situation to recover.: \textit{4}
        \item The removal has no effect because it is ignored by the people in the discussion.: \textit{7}
        \item The removal has no effect because the discussion has already ended.: \textit{10}
    \end{itemize}
\end{enumerate}
\end{adjustwidth}
\begin{enumerate}[start=7,label={\bfseries Q\arabic*:}]
    \item Have you ever made a comment on r/changemyview that you later regretted because in hindsight it could be perceived as offensive or uncivil?
    \begin{itemize}
        \item Never.: \textit{15}
        \item Yes, and the moderators removed it.: \textit{4}
        \item Yes, and I later removed it myself.: \textit{19}
        \item Yes, and it was never removed.: \textit{9}
    \end{itemize}
    \item Which of the following statements about r/changemyview's enforcement of Rule 2 do you agree with? (Check all that apply)
    \begin{itemize}
        \item I am satisfied with the existing enforcement.: \textit{28}
        \item The existing enforcement is too much (comments often get removed that didn't deserve it).: \textit{7}
        \item The existing enforcement is not enough (comments that deserve to be removed often aren't).: \textit{9}
        \item The existing enforcement is biased.: \textit{6}
        \item It is too hard to get a bad enforcement decision overturned.: \textit{4}
        \item Enforcement needs to be more transparent.: \textit{16}
    \end{itemize}
    \item Are there any other things you wish r/changemyview did differently in enforcing Rule 2? \textit{[Free response (See Appendix \ref{appendix:freeresponse} for sampled answers)]}
\end{enumerate}

\noindent\underline{Part 3: Forecasting incivility}

The following questions will ask about your personal intuitions about when incivility occurs in discussions. We emphasize that you should answer these questions from the perspective of your own intuitions, \textbf{without} the help of ConvoWizard.

\begin{enumerate}[start=10,label={\bfseries Q\arabic*:}]
    \item Can you personally tell when discussions are at risk of turning uncivil (that is, may later lead to comments that will violate Rule 2)?
    \begin{itemize}
        \item I cannot tell.: \textit{0}
        \item I can tell in some cases.: \textit{19}
        \item I can tell in many cases.: \textit{18}
        \item I can tell in most cases.: \textit{10}
    \end{itemize}
\end{enumerate}

\begin{adjustwidth}{0.75in}{}
\textit{Show the following question(s) if ``I cannot tell'' was NOT selected in \textbf{Q10} (47 participants):}
\begin{enumerate}[start=11,label={\bfseries Q\arabic*:}]
    \item Briefly explain how you can tell if a discussion is at risk of turning uncivil. \textit{[Free response]}
    \item If you think a discussion is at risk of turning uncivil, does this make you more willing or less willing to participate?
    \begin{itemize}
        \item More willing: \textit{2}
        \item Less willing: \textit{29}
        \item No effect: \textit{16}
    \end{itemize}
    \item If you think a discussion is at risk of turning uncivil and you are participating, does this affect how you phrase your comments?
    \begin{itemize}
        \item Yes: \textit{36}
        \item No: \textit{11}
    \end{itemize}
\end{enumerate}
\begin{adjustwidth}{0.75in}{}
\textit{Show the following question(s) if ``Yes'' was selected in \textbf{Q13} (36 participants):}
\begin{enumerate}[start=14,label={\bfseries Q\arabic*:}]
    \item How does the phrasing you use in your comments change when you think the discussion is at risk of turning uncivil? Select all that apply:
    \begin{itemize}
        \item I use more polite language.: \textit{19}
        \item I use fewer swear words.: \textit{2}
        \item I use more formal language.: \textit{17}
        \item I use more casual language.: \textit{4}
        \item I use more objective language (that is, I try to frame my comment in terms of facts and data).: \textit{24}
        \item I use more subjective language (that is, I try to frame my comment in terms of personal feelings and opinions).: \textit{4}
        \item I ask more questions.: \textit{18}
        \item I write a shorter comment.: \textit{10}
        \item I write a longer comment.: \textit{11}
        \item Other (please describe):: \textit{9}
    \end{itemize}
\end{enumerate}
\end{adjustwidth}
\end{adjustwidth}

\noindent\underline{Part 4: Experience with ConvoWizard: Context Summary Feedback}

The following questions will ask about your experience with the Context summary feedback feature of ConvoWizard. This is referring to the top box that gave a summary of how likely the preexisting discussion was to turn uncivil before you joined (see the highlighted part of the screenshot below): \textit{[Screenshot of ConvoWizard interface with Context Summary box highlighted]}

\begin{enumerate}[start=15,label={\bfseries Q\arabic*:}]
    \item Do you remember seeing the text and/or color of the context summary box change (indicating that the discussion might be getting tense)?
    \begin{itemize}
        \item Yes: \textit{38}
        \item No: \textit{9}
    \end{itemize}
\end{enumerate}
\begin{adjustwidth}{0.75in}{}
\textit{Show the following question(s) if ``Yes'' was selected in \textbf{Q15} (38 participants):}
\begin{enumerate}[start=16,label={\bfseries Q\arabic*:}]
    \item Thinking specifically of times when you saw the text and/or color of the context summary box change, did the context summary feedback ever...
    \begin{enumerate}[a)]
        \item ...help you avoid a fight or confrontation?
        \begin{itemize}
            \item Yes: \textit{19}
            \item No: \textit{19}
        \end{itemize}
        \item ...affect whether you decided to post a reply?
        \begin{itemize}
            \item Yes: \textit{20}
            \item No: \textit{18}
        \end{itemize}
        \item ...affect what you said in your reply, if you posted one?
        \begin{itemize}
            \item Yes: \textit{26}
            \item No: \textit{12}
        \end{itemize}
    \end{enumerate}
\end{enumerate}
\begin{adjustwidth}{0.75in}{}
\textit{Show the following question(s) if ``Yes'' was selected in \textbf{Q16c} (26 participants):}
\begin{enumerate}[start=17,label={\bfseries Q\arabic*:}]
    \item Thinking specifically of times when you saw the text and/or color of the context summary box change, how did the context summary feedback affect what you said in your reply? Select all that apply:
    \begin{itemize}
        \item I used more polite language.: \textit{17}
        \item I used fewer swear words.: \textit{1}
        \item I used more formal language.: \textit{7}
        \item I used more casual language.: \textit{3}
        \item I used more objective language (that is, I try to frame my comment in terms of facts and data).: \textit{9}
        \item I used more subjective language (that is, I try to frame my comment in terms of personal feelings and experiences).: \textit{2}
        \item I asked more questions.: \textit{9}
        \item I wrote a shorter comment.: \textit{8}
        \item I wrote a longer comment.: \textit{2}
        \item Other (please describe): \textit{4}
    \end{itemize}
\end{enumerate}
\end{adjustwidth}
\begin{enumerate}[start=18,label={\bfseries Q\arabic*:}]
    \item Overall, how useful was the context summary feedback?
    \begin{itemize}
        \item Not at all useful: \textit{8}
        \item Somewhat useful: \textit{19}
        \item Quite useful: \textit{10}
        \item Very useful: \textit{1}
    \end{itemize}
\end{enumerate}
\end{adjustwidth}
\begin{enumerate}[start=19,label={\bfseries Q\arabic*:}]
    \item Do you think ConvoWizard is better or worse than you at telling whether a discussion might be getting tense?
    \begin{itemize}
        \item Much better: \textit{2}
        \item Somewhat better: \textit{7}
        \item About the same: \textit{16}
        \item Somewhat worse: \textit{15}
        \item Much worse: \textit{7}
    \end{itemize}
\end{enumerate}
\begin{adjustwidth}{0.75in}{}
\textit{Show the following question(s) if ``Much better'' or ``Somewhat better'' was selected in \textbf{Q19} (9 participants):}
\begin{enumerate}[start=20,label={\bfseries Q\arabic*:}]
    \item Why do you think ConvoWizard is better than you at telling whether a discussion might be getting tense? \textit{[Free response]}
\end{enumerate}
\end{adjustwidth}
\begin{adjustwidth}{0.75in}{}
\textit{Show the following question(s) if ``Much worse'' or ``Somewhat worse'' was selected in \textbf{Q19} (22 participants):}
\begin{enumerate}[start=21,label={\bfseries Q\arabic*:}]
    \item Why do you think ConvoWizard is worse than you at telling whether a discussion might be getting tense? \textit{[Free response]}
\end{enumerate}
\end{adjustwidth}
\begin{enumerate}[start=22,label={\bfseries Q\arabic*:}]
    \item For which of the following reasons, if any, did you ever disagree with the context summary feedback? ``Disagree'' means that you intuitively felt the feedback was wrong, or you would have made a different judgment call. Rate how often each potential disagreement occurred on a scale from ``Never'' to ``Very often''.
    \begin{enumerate}[a)]
        \item ConvoWizard said a discussion looked tense even though it wasn't
        \begin{itemize}
            \item Never: \textit{6}
            \item Rarely: \textit{12}
            \item Sometimes: \textit{21}
            \item Often: \textit{7}
            \item Very often: \textit{1}
        \end{itemize}
        \item ConvoWizard did not say a discussion was tense even though it clearly was.
        \begin{itemize}
            \item Never: \textit{14}
            \item Rarely: \textit{18}
            \item Sometimes: \textit{14}
            \item Often: \textit{0}
            \item Very often: \textit{1}
        \end{itemize}
        \item ConvoWizard's estimated degree of tension was incorrect (for example, a discussion was marked as ``somewhat'' tense when it was actually extremely tense).
        \begin{itemize}
            \item Never: \textit{14}
            \item Rarely: \textit{13}
            \item Sometimes: \textit{13}
            \item Often: \textit{5}
            \item Very often: \textit{2}
        \end{itemize}
        \item ConvoWizard's context summary feedback seemed to be biased.
        \begin{itemize}
            \item Never: \textit{27}
            \item Rarely: \textit{13}
            \item Sometimes: \textit{6}
            \item Often: \textit{1}
            \item Very often: \textit{0}
        \end{itemize}
    \end{enumerate}
    \item Are there any other reasons not listed above that you disagreed with the context summary feedback? (You can also use this space to elaborate on your answers to the previous question). \textit{[Free response]}
\end{enumerate}

\noindent\underline{Part 5: Experience with ConvoWizard: Reply Summary Feedback}

The following questions will ask about your experience with the Reply summary feedback feature of ConvoWizard. This is referring to the bottom box that gave a summary of how the reply you were drafting could affect the tension in the discussion if it was posted (see the highlighted part of the screenshot below): \textit{[Screenshot of ConvoWizard with the Reply Summary box highlighted]}

\begin{enumerate}[start=24,label={\bfseries Q\arabic*:}]
    \item Do you remember seeing the text and/or color of the reply summary box change (indicating potential increase or decrease in tension)?
    \begin{itemize}
        \item Yes: \textit{35}
        \item No: \textit{12}
    \end{itemize}
\end{enumerate}
\begin{adjustwidth}{0.75in}{}
\textit{Show the following question(s) if ``Yes'' was selected in \textbf{Q24} (35 participants):}
\begin{enumerate}[start=25,label={\bfseries Q\arabic*:}]
    \item Thinking specifically of times when you saw the text and/or color of the reply summary box change, did the reply summary feedback ever...
    \begin{enumerate}[a)]
        \item ...help you avoid a fight or confrontation?
        \begin{itemize}
            \item Yes: \textit{19}
            \item No: \textit{16}
        \end{itemize}
        \item ...stop you from posting something you might have regretted later?
        \begin{itemize}
            \item Yes: \textit{19}
            \item No: \textit{16}
        \end{itemize}
        \item ...affect whether you decided to eventually post your draft reply?
        \begin{itemize}
            \item Yes: \textit{21}
            \item No: \textit{14}
        \end{itemize}
        \item ...affect what you said in the reply you ended up posting, if you posted one?
        \begin{itemize}
            \item Yes: \textit{25}
            \item No: \textit{10}
        \end{itemize}
    \end{enumerate}
\end{enumerate}
\begin{adjustwidth}{0.75in}{}
\textit{Show the following question(s) if ``Yes'' was selected in \textbf{Q25c} (25 participants):}
\begin{enumerate}[start=26,label={\bfseries Q\arabic*:}]
    \item Thinking specifically of times when you saw the text and/or color of the reply summary box change to indicate an increase in tension (i.e. a reddish color), how did the reply summary feedback change what you said in your reply? Select all that apply:
    \begin{itemize}
        \item N/A (I have never seen an increase in tension).: \textit{0}
        \item I used more polite language.: \textit{17}
        \item I used fewer swear words.: \textit{2}
        \item I used more formal language.: \textit{12}
        \item I used more casual language.: \textit{5}
        \item I used more objective language (that is, I try to frame my comment in terms of facts and data).: \textit{11}
        \item I used more subjective language (that is, I try to frame my comment in terms of personal feelings and experiences).: \textit{1}
        \item I asked more questions.: \textit{8}
        \item I wrote a shorter comment.: \textit{4}
        \item I wrote a longer comment.: \textit{4}
        \item Other (please describe): \textit{3}
    \end{itemize}
\end{enumerate}
\end{adjustwidth}
\begin{enumerate}[start=27,label={\bfseries Q\arabic*:}]
    \item Overall, how useful was the reply summary feedback?
    \begin{itemize}
        \item Not at all useful: \textit{8}
        \item Somewhat useful: \textit{17}
        \item Quite useful: \textit{10}
        \item Very useful: \textit{0}
    \end{itemize}
\end{enumerate}
\end{adjustwidth}
\begin{enumerate}[start=28,label={\bfseries Q\arabic*:}]
    \item Do you think ConvoWizard is better or worse than you at telling whether a draft reply might increase tension in the discussion?
    \begin{itemize}
        \item Much better: \textit{1}
        \item Somewhat better: \textit{8}
        \item About the same: \textit{23}
        \item Somewhat worse: \textit{12}
        \item Much worse: \textit{3}
    \end{itemize}
\end{enumerate}
\begin{adjustwidth}{0.75in}{}
\textit{Show the following question(s) if ``Much better'' or ``Somewhat better'' was selected in \textbf{Q28} (9 participants):}
\begin{enumerate}[start=29,label={\bfseries Q\arabic*:}]
    \item Why do you think ConvoWizard is better than you at telling whether a draft reply might increase tension in the discussion? \textit{[Free response]}
\end{enumerate}
\end{adjustwidth}
\begin{adjustwidth}{0.75in}{}
\textit{Show the following question(s) if ``Much worse'' or ``Somewhat worse'' was selected in \textbf{Q28} (15 participants):}
\begin{enumerate}[start=30,label={\bfseries Q\arabic*:}]
    \item Why do you think ConvoWizard is worse than you at telling whether a draft reply might increase tension in the discussion? \textit{[Free response]}
\end{enumerate}
\end{adjustwidth}
\begin{enumerate}[start=31,label={\bfseries Q\arabic*:}]
    \item For which of the following reasons, if any, did you ever disagree with the reply summary feedback? ``Disagree'' means that you intuitively felt the feedback was wrong, or you would have made a different judgment call. Rate how often each potential disagreement occurred on a scale from ``Never'' to ``Very often''.
    \begin{enumerate}[a)]
        \item ConvoWizard said my reply would increase tension even though it clearly wouldn't.
        \begin{itemize}
            \item Never: \textit{9}
            \item Rarely: \textit{9}
            \item Sometimes: \textit{20}
            \item Often: \textit{6}
            \item Very often: \textit{3}
        \end{itemize}
        \item ConvoWizard did not say my reply would increase tension even though it clearly would.
        \begin{itemize}
            \item Never: \textit{20}
            \item Rarely: \textit{11}
            \item Sometimes: \textit{13}
            \item Often: \textit{2}
            \item Very often: \textit{1}
        \end{itemize}
        \item Changing the text of my draft did not seem to change what ConvoWizard said.
        \begin{itemize}
            \item Never: \textit{13}
            \item Rarely: \textit{12}
            \item Sometimes: \textit{15}
            \item Often: \textit{6}
            \item Very often: \textit{1}
        \end{itemize}
        \item A minor/trivial change to the text of my draft changed what ConvoWizard said.
        \begin{itemize}
            \item Never: \textit{11}
            \item Rarely: \textit{7}
            \item Sometimes: \textit{15}
            \item Often: \textit{9}
            \item Very often: \textit{5}
        \end{itemize}
        \item ConvoWizard's reply summary feedback seemed to be biased.
        \begin{itemize}
            \item Never: \textit{27}
            \item Rarely: \textit{12}
            \item Sometimes: \textit{8}
            \item Often: \textit{0}
            \item Very often: \textit{0}
        \end{itemize}
    \end{enumerate}
    \item Are there any other reasons not listed above that you disagreed with the reply summary feedback? (You can also use this space to elaborate on your answers to the previous question). \textit{[Free response]}
\end{enumerate}

\noindent\underline{Part 6: Overall impressions}

The following questions ask about your overall impressions of ConvoWizard, accounting for all its features.
\begin{enumerate}[start=33,label={\bfseries Q\arabic*:}]
    \item Between the context summary feedback and reply summary feedback, which did you find more helpful?
    \begin{itemize}
        \item Context summary: \textit{7}
        \item Reply summary: \textit{18}
        \item Both were equally helpful: \textit{8}
        \item Both were equally unhelpful: \textit{14}
    \end{itemize}
    \item If ConvoWizard were to be publicly released and worked on all versions of Reddit (including new Reddit and mobile), how likely would you be to use it as part of your usual r/changemyiew participation?
    \begin{itemize}
        \item I would definitely not use it.: \textit{8}
        \item I might try it.: \textit{18}
        \item I would probably try it.: \textit{13}
        \item I would definitely use it.: \textit{3}
        \item I would definitely use it, and recommend it to others.: \textit{5}
    \end{itemize}
    \item If ConvoWizard were to be publicly released and many members of r/changemyview used it, do you think this would improve or harm overall discussion quality?
    \begin{itemize}
        \item It would improve discussion quality: \textit{30}
        \item It would harm discussion quality: \textit{1}
        \item It would have little to no effect: \textit{16}
    \end{itemize}
    \item Which would you prefer to use: ConvoWizard (which predicts whether a discussion / comment might lead to uncivil behavior in the future), or a tool that detects whether a discussion/comment is already uncivil?
    \begin{itemize}
        \item I would prefer ConvoWizard.: \textit{25}
        \item I would prefer the tool that detects already existing incivility.: \textit{5}
        \item I would use both.: \textit{7}
        \item I would use neither.: \textit{10}
        \item I cannot tell the difference.: \textit{0}
    \end{itemize}
    \item Which of the following improvements would be most important to you in deciding to use or recommend ConvoWizard? (Select up to 3)
    \begin{itemize}
        \item Correctly identifying more of the tense discussions or draft replies.: \textit{18}
        \item Giving fewer false alerts on harmless discussions or replies.: \textit{17}
        \item Better user interface and integration with the Reddit webpage.: \textit{14}
        \item More consistent behavior.: \textit{7}
        \item More transparency (i.e., explanations of why ConvoWizard marked a discussion / comment as tense).: \textit{23}
        \item More concrete suggestions on how to decrease tension: \textit{14}
        \item Availability on other platforms (new Reddit, mobile app, etc.).: \textit{13}
        \item Other (please describe): \textit{6}
    \end{itemize}
    \item Did you encounter any technical issues while using ConvoWizard?
    \begin{itemize}
        \item Yes (please describe): \textit{0}
        \item No: \textit{37}
    \end{itemize}
    \item Would you be interested in continuing to test ConvoWizard, assuming we extend the testing period?  This is entirely optional and the answer to this question will not affect your receipt of the \$20 gift card for the testing period you just finished.
    \begin{itemize}
        \item Yes: \textit{33}
        \item No: \textit{14}
    \end{itemize}
    \item In the case the results of this study will be published in a scientific article, would you be OK with us anonymously quoting your answers you provided in this survey?  We will not disclose your Reddit username (or any other identity).
    \begin{itemize}
        \item Yes: \textit{44}
        \item No: \textit{3}
    \end{itemize}
\end{enumerate}

\section{Sampled Free Responses}
\label{appendix:freeresponse}

For each free response question, we have randomly sampled three responses to be shown as examples (unless there were fewer than three total responses, for optional / conditional questions).
\vspace{\baselineskip}

\noindent Are there any other things you wish r/changemyview did differently in enforcing Rule 2?
\begin{itemize}
\item Being consistent. CMV removes certain comments, but far after the conversation dissolves into insults and hostility.
\item There are clearly a large bias present in the subreddit, particularly on topics that if you are not going along with what is the `popular' thing then you get downvoted, or just insulted.
\item No, the moderators are great with enforcement.
\end{itemize}

\noindent Briefly explain how you can tell if a discussion is at risk of turning uncivil.
\begin{itemize}
\item Just a feeling that some people are starting more hostile than others.
\item If the conversation starts getting personal, attacking personal credentials or identity instead of the problem. 
\item The easiest way is to analyze the phrasing. Stern, short phrases, completely contradicting the other person's viewpoint might come off as hostile and aggressive, causing a defensive reaction that might turn into an uncivil discussion.
\end{itemize}

\noindent How does the phrasing you use in your comments change when you think the discussion is at risk of turning uncivil? Select all that apply: - Other (please describe): 
\begin{itemize}
\item I give minor concessions to points they have made
\item try to explain why you see the way you do and what makes you disagree with them.
\item Pretty much all of the above to some degree. I never want to offend anyone. And I try not to be offended. Swearing just turns things instantly uncivil. 
\end{itemize}

\noindent Thinking specifically of times when you saw the text and/or color of the context summary box change, how did the context summary feedback affect what you said in your reply? Select all that apply: - Other (please describe) 
\begin{itemize}
\item All of the above again. I thought of better words I could usem maybe words that don't sound like I may be trying to provoke a uncivil response. I tried longer comments as I am not good at summarizing things in short comments. I enjoyed having this tool to help me see things I may not realize I am posting. 
\item I kept rewording my reply until it stopped showing up orange.  It usually led to less effective replies that, in retrospect, were too wishy-washy to change anyone's view.
\item I tended to avoid certain key words that I felt the program picked up on whether or not I was being confrontational. The word ``you'' or any words with negative connotations could be altered without changing the meat of my messages. 
\end{itemize}

\noindent Why do you think ConvoWizard is better than you at telling whether a discussion might be getting tense?
\begin{itemize}
\item Often times if the color changed I would reread what I was saying and see if the response maybe cam off the wrong way. Helping me then to reword it. 
\item It seems to be able to sense strong emotions, but it doesn't seem to understand pathos arguments.
\item It's hard in the moment when reading a divisive comment to objectively recognize where the conversation is going
\end{itemize}

\noindent Why do you think ConvoWizard is worse than you at telling whether a discussion might be getting tense?
\begin{itemize}
\item I tried to test its capabilities. In my experience, direct insults do not necessarily alert the program of anything being wrong. The comment has to be sufficiently long for it to usually detect possible cases of uncivility rising. It also seems a little too sensitive, sometimes a comment that was meant to be stern alerts ConvoWizard.
\item ConvoWizard seemed to be based off of specific words being in the conversation at all? Discussion on the r-slur were always red, because the word set ConvoWizard off. Quoting other people's tens dialog also seemed to affect that Wizard just as much as saying it myself, but quoting other people's dialogue is just required to have the discussion. 
\item It said everything was at risk of getting tense
\end{itemize}

\noindent Are there any other reasons not listed above that you disagreed with the context summary feedback? (You can also use this space to elaborate on your answers to the previous question).
\begin{itemize}
\item Yes, sometimes the conversation was becoming tense and ConvoWizard didn't notice it. 
\item Frankly, I just didn't encounter many tense arguments. I was impressed with the tool's sentiment analysis, but I don't have any evidence that it could identify tension that I or most other commenters would fail to identify.
\item It got obvious things right but didn't seem to work well on the fringe cases. 
\end{itemize}

\noindent Thinking specifically of times when you saw the text and/or color of the reply summary box change to indicate an increase in tension (i.e. a reddish color), how did the reply summary feedback change what you said in your reply? Select all that apply: - Other (please describe) 
\begin{itemize}
\item See previous answer.  I reworded it.  Looking back, I disagree with my rewording and think my posts became less likely to earn a delta.
\item I'm not entirely sure what changed but that I did
\end{itemize}

\noindent Why do you think ConvoWizard is better than you at telling whether a draft reply might increase tension in the discussion?
\begin{itemize}
\item It's easy to pick up the read tense but I'm not always sure when what I'm going to say will make things better or worse.
\item Certain verbiage that I typically used the wizard pointed out and I adjusted the verbiage. 
\item I don't often care about increasing tension. My objective is generally the discussion, not whether I sound polite or not. ConvoWizard sort of reminds me that I should use maybe different language.
\end{itemize}

\noindent Why do you think ConvoWizard is worse than you at telling whether a draft reply might increase tension in the discussion?
\begin{itemize}
\item Sometimes it seemed to think a very innocuous response would escalate tension when I found that unlikely. 
\item I just don't think the extension works very well. It must be a technical issue. 
\item It reacted to obvious stimuli but didn't work well with sarcasm or curtness, which are often the first signs that a conversation is becoming tense. 
\end{itemize}

\noindent Are there any other reasons not listed above that you disagreed with the reply summary feedback? (You can also use this space to elaborate on your answers to the previous question).
\begin{itemize}
\item No. 
\item Just as before, when quoting someone else's text, ConvoWizard treated it as if the person themselves was saying it. This misrepresents the discussion. 
\item Primarily that it seemed to say everything was in danger of tension
\end{itemize}

\noindent Which of the following improvements would be most important to you in deciding to use or recommend ConvoWizard? (Select up to 3) - Other (please describe) 
\begin{itemize}
\item Firefox please. 
\item I'm perfectly able to tell if people are getting `tense'. I don't need software to tell me. 
\item Honestly, the biggest issue is me. I almost always knew when a conversation was getting uncivil, but was going to post regardless. The wizard rarely shamed me into not posting (although it did work occasionally! which was surprising). Granted, i do use reddit as an outlet to vent/argue, so i wasn't really trying to avoid being uncivil.  If it doesn't change my behavior, it doesn't do much to warn me something is uncivil  
\end{itemize}

\noindent Did you encounter any technical issues while using ConvoWizard? - Yes (please describe) 
\begin{itemize}
\item It occasionally would stop returning a result mid-reply, or not really return a result at all.
\item My anti-virus flagged it once.
\item text boxes that are light gray on white. v hard to read.
\end{itemize}

\noindent Is there any additional feedback you would like to provide that was not already covered, or anything in particular that you liked or disliked?
\begin{itemize}
\item I would love to see this as a feature on Reddit in general. It could really help things. Though if it would change how people act is unseen. 
\item It was hard to use, since I had to use old Reddit. It made me use it less often
\item no
\end{itemize}

\end{document}